\documentclass[12pt]{article}
\usepackage[total={13.5cm,18.3cm},centering]{geometry}
\usepackage{amsfonts, amsmath}

\newcommand{\be}{\begin{equation}} \newcommand{\ee}{\end{equation}}

\title{\bf
The Quantum and Statistical Theory of Early Universe and Its
Possible Applications to Cosmology} 

\author{Alex.E.Shalyt-Margolin\hspace{1.5mm}\thanks
{Phone (+375) 172 883438; e-mail: alexm@hep.by; a.shalyt@mail.ru}
\thanks{Fax (+375) 172 926075}
}
\date{}
\begin{document}

\maketitle
 {\footnotesize\noindent Laboratory of the Quantum Field Theory,
National Center of Particles and High Energy Physics, Bogdanovich
Str. 153, Minsk 220040, Belarus\\ {{\footnotesize
}}

\rm\normalsize 
\begin{abstract}

The subject of this study is Quantum and Statistical Mechanics of
the Early Universe.In it  a new approach to investigation  of
these two theories - density matrix deformation - is proposed. The
distinguishing feature of the proposed approach as compared to the
previous ones is the fact that here the density matrix is
subjected to deformation, rather than commutators or (that is the
same) Heisenberg's Algebra. The deformation is understood as an
extension of a particular theory by inclusion of one or several
additional parameters in such a way that the initial theory
appears in the limiting transition.Some consequences of this
approach for unitarity problem in Early Universe,black hole
entropy,information paradox problem,for symmetry restoration in
the associated deformed field model with scalar field are
proposed.
\end{abstract}

\section{Introduction}
At the present time it is clear that a quantum theory of the Early
Universe (Planck's energies  and energies close to the Planck's)
should be distinguished from the corresponding theory at the
well-known energies. Specifically, a quantum theory of the Early
Universe should include a minimal length. In the last few years the
Early Universe has aroused considerable interest of the researchers.
This may be caused by several facts. First, a Big Bang theory is
presently well grounded and has established experimental status.
Second, acknowledged success of the inflation model and its
interface with high-energy physics. Third, various approaches to
topical problems of the fundamental physics, specifically to the
problem of divergence in a quantum theory or singularity in the
General Relativity Theory, in some or other way lead to the problem
of quantum-gravitational effects and their adequate description. And
all the above aspects are related to the Early Universe. Because of
this, investigation of the Early Universe is of particular
importance. The Early Universe is understood as a Universe at the
first Planck's moments following the Big Bang when energies and
scales were on the order of Planck's.

In this paper a new approach to investigation of Quantum and
Statistical Mechanics of the Early Universe  - density matrix
deformation - is proposed. The deformation is understood as an
extension of a particular theory by inclusion of one or several
additional parameters in such a way that the initial theory appears
in the limiting transition. The most clear example is QM being a
deformation of Classical Mechanics (CM). The parameter of
deformation in this case is the Planck's constant $\hbar$. When
$\hbar\rightarrow 0$ QM goes to Classical Mechanics.

In the first part of this article Quantum Mechanics of the Early
Universe is treated as a Quantum Mechanics with Fundamental Length.
This becomes possible since different approaches to quantum
gravitation exhibited in the Early Universe unavoidably involve the
notion of fundamental length on the order of Planck's (see \cite{r1}
and the references). Also this is possible due to the involvement in
this theory of the Generalized Uncertainty Relations. And Quantum
Mechanics with Fundamental Length is obtained as a deformation of
Quantum Mechanics. The distinguishing feature of the proposed
approach as compared to the previous ones is the fact that here the
density matrix is subjected to deformation, rather than commutators
or (that is the same) Heisenberg's Algebra. In this chapter the
density matrix obtained by deformation of the quantum-mechanical one
is referred to as a density pro-matrix. Within our approach two main
features of Quantum Mechanics are conserved: the probabilistic
interpretation of the theory and the well-known measuring procedure
associated with this interpretation. The proposed approach allows
for description of dynamics, in particular, the explicit form of
deformed Liouville equation and the deformed Shrodinger's picture.
Some implications of the obtained results are discussed including
the singularity problem, hypothesis of cosmic censorship, possible
improvement of the definition for statistical entropy. It is shown
that owing to the obtained results one is enabled to deduce in a
simple and natural way the Bekenstein-Hawking formula for black hole
entropy in a semiclassical approximation. In the second part of the
chapter it is demonstrated that Statistical Mechanics of the Early
Universe is a deformation of the conventional Statistical Mechanics.
The statistical-mechanics deformation is constructed by analogy to
the earlier quantum mechanical results. As previously, the primary
object is a density matrix, but now the statistical one. The
obtained deformed object is referred to as a statistical density
pro-matrix. This object is explicitly described, and it is
demonstrated that there is a complete analogy in the construction
and properties of quantum-mechanics and statistical density matrices
at Plank scale (i.e. density pro-matrices). It is shown that an
ordinary statistical density matrix occurs in the low-temperature
limit at temperatures much lower than the Plank's. The associated
deformation of a canonical Gibbs distribution is given explicitly.
Also consideration is being given to rigorous substantiation of the
Generalized Uncertainty Relations as applied in thermodynamics. And
in the third part of the chapter the results obtained are applied to
solution of the Information Paradox (Hawking) Problem. It is
demonstrated that involvement of black holes in the suggested
approach in the end twice causes nonunitary transitions resulting in
the unitarity. In parallel this problem is considered in other
terms: entropy density, Heisenberg algebra deformation terms,
respective deformations of Statistical Mechanics, - all showing the
identity of the basic results. From this an explicit solution for
Hawking's  Informaion paradox has been derived. Besides, it is shown
that owing to the proposed approach a new small parameter is derived
in physics, the principal features of which are its dimensionless
character and its association with all the fundamental constants.
Then a possible explanation to J.Bekenstein's \cite{bek1} problem
about high entropy of the Planck' black-hole remainders is given in
terms of  the entropy density matix, and also, proceeding from the
results of R.Bousso \cite{bou2}, a hypothesis concerning possible
inference from the holographic principle for strong gravitational
field is set up. In the last part of the paper it is shown that on
the basis of the above parameter the Universe may be considered as
nonuniform lattice in the finite-dimensional hypercube. Besides,
possible applications of the results are proposed. The case of
spontaneous symmetry breaking and restoration for simple Lagrangian
with scalar field is considered on this lattice in detail together
with a number of inferences for cosmology.

 This paper is a revised and extended version of the previous work
\cite{shalyt20}.The principal findings presented in this paper
have been published in a series of works
\cite{shalyt5},\cite{shalyt7},\cite{shalyt13},\cite{shalyt13s},
\cite{shalyt14},\cite{shalyt15}. These findings were given in greater
detail in \cite{shalyt16},\cite{shalyt17},
\cite{shalyt18},\cite{shalyt19}.

\section{Fundamental Length and Density Matrix}

Using different approaches (String Theory \cite{r2}, Gravitation
\cite{r3}, etc.), the authors of numerous papers issued over the
last 14-15 years have pointed out that Heisenberg's Uncertainty
Relations should be modified. Specifically, a high energy
correction has to appear
\begin{equation}\label{U2}
\triangle x\geq\frac{\hbar}{\triangle p}+\alpha^{\prime}
L_{p}^2\frac{\triangle p}{\hbar}.
\end{equation}
\noindent Here $L_{p}$ is the Planck's length:
$L_{p}=\sqrt\frac{G\hbar}{c^3}\simeq1,6\;10^{-35}\;m$ and
 $\alpha^{\prime} > 0$ is a constant. In \cite{r3} it was shown
that this constant may be chosen equal to 1. However, here we will
use $\alpha^{\prime}$ as an arbitrary constant without giving it any
definite value. Equation (\ref{U2})  is identified as the
Generalized Uncertainty Relations in Quantum Mechanics.

The inequality (\ref{U2}) is quadratic in $\triangle p$:
\begin{equation}\label{U3}
\alpha^{\prime} L_{p}^2({\triangle p})^2-\hbar \triangle x
\triangle p+ \hbar^2 \leq0,
\end{equation}
from whence the fundamental length is
\begin{equation}\label{U4}
\triangle x_{min}=2\sqrt\alpha^{\prime} L_{p}.
\end{equation}
Since in what follows we proceed only from the existence of
fundamental length, it should be noted that this fact was
established apart from GUR as well. For instance, from an ideal
experiment associated with Gravitational Field and Quantum
Mechanics a lower bound on minimal length was obtained in
\cite{r7}, \cite{r8} and  improved in \cite{r9} without using GUR
to an estimate of the form $\sim L_{p}$. As reviewed previously in
\cite{r1}, the fundamental length appears quite naturally at
Planck scale, being related to the quantum-gravitational
effects.\noindent Let us  consider equation (\ref{U4}) in some
detail.  Squaring both its sides, we obtain
\begin{equation}\label{U5}
(\overline{\Delta\widehat{X}^{2}})\geq 4\alpha^{\prime} L_{p}^{2},
\end{equation}
Or in terms of density matrix
\begin{equation}\label{U6}
Sp[(\rho \widehat{X}^2)-Sp^2(\rho \widehat{X}) ]\geq
4\alpha^{\prime} L_{p}^{2 }=l^{2}_{min}>0,
\end{equation}
where $\widehat{X}$ is the coordinate operator. Expression
(\ref{U6}) gives the measuring rule used in QM.As distinct from
QM,however, in the are considered here the right-hand side of
(\ref{U6}) can not be brought arbitrary close to zero as it is
limited by $l^{2}_{min}>0$, where because of GUR $l_{min} \sim
L_{p}$.

Apparently, this may be due to the fact that QMFL is unitary
non-equivalent to  QM. Actually, in QM the left-hand side of
(\ref{U6}) can be chosen arbitrary close to zero, whereas in QMFL
this is impossible. But if two theories are unitary equivalent then
the form of their spurs should be retained. Besides, a more
important aspect is contributing to unitary non-equivalence of these
two theories: QMFL contains three fundamental constants (independent
parameters) $G$, $c$ and $\hbar$, whereas QM contains only one:
$\hbar$. Within an inflationary model (see \cite{r10}), QM is the
low-energy limit of QMFL (QMFL turns to QM) for the expansion of the
Universe.This is identical for all cases of transition from Planck's
energies to the normal ones \cite{r1}. In special case of using GUR,
the second term in the right-hand side of (\ref{U2}) vanishes and
GUR turn to UR \cite{r6}. A natural way for studying QMFL is to
consider this theory as a deformation of QM, turning to QM at the
low energy limit (during expansion of the Universe after the Big
Bang). We will consider precisely this option.In this paper, unlike
the works of other authors (e.g. see \cite{r5}) the density matrix
is deformed rather than commutators, whereas the fundamental
fundamental quantum-mechanical measuring rule (\ref{U6}) is left
without changes. Here the following question may be formulated: how
should be deformed density matrix conserving quantum-mechanical
measuring rules in order to obtain self-consistent measuring
procedure in QMFL? To answer this question, we use the R-procedure.
First consider R-procedure both at the Planck's and  low-energy
scales. At the Planck's scale $a \approx il_{min}$ or $a \sim
iL_{p}$, where $i$ is a small quantity. Further $a$ tends to
infinity and we obtain for density matrix
\cite{shalyt1}-\cite{shalyt5}:
 $$Sp[\rho a^{2}]-Sp[\rho
a]Sp[\rho a] \simeq l^{2}_{min}\;\; or\;\; Sp[\rho]-Sp^{2}[\rho]
\simeq l^{2}_{min}/a^{2}.$$

 Therefore:

 \begin{enumerate}
 \item When $a < \infty$, $Sp[\rho] =
Sp[\rho(a)]$ and
 $Sp[\rho]-Sp^{2}[\rho]>0$. Then  \newline $Sp[\rho]<1$
 that corresponds to the QMFL case.
\item When $a = \infty$, $Sp[\rho]$ does not depend on $a$ and
$Sp[\rho]-Sp^{2}[\rho]\rightarrow 0$. Then  $Sp[\rho]=1$ that
corresponds to the QM case.
\end{enumerate}
Interesting,how should be  interpreted 1 and 2 ? Does  the above
analysis agree with the main result from \cite{r30} \footnote
{``\dots there cannot be any physical state which is a position
eigenstate since an eigenstate would of course have zero uncertainty
in position''}? Note the agreement is well. Indeed, any time when
the state vector reduction (R-procedure) place in QM always an
eigenstate (value) is chosen exactly. In other words, the
probability is equal to 1. As it was pointed out in statement 1, the
situation changes when we consider QMFL: it is impossible to measure
coordinates exactly,they never will be absolutely reliable. In all
cases  we obtain a probability less than 1 ($Sp[\rho]=p<1$). In
other words, any R-procedure in QMFL leads to an eigenvalue, but
only with a probability less than 1. This probability is as near to
1 as far the difference between measuring scale $a$ and $l_{min}$ is
growing, or in other words, the second term in (\ref{U2}) becomes
insignificant and we turn to QM. Here there is no  contradiction
with \cite{r30}. In QMFL there are no exact coordinate eigenstates
(values) as well as there are no pure states. In this paper we
consider not the operator properties in QMFL as it was done in
\cite{r30} but density-matrix properties.

The  properties of density matrix in QMFL and QM have to be
different. The only reasoning in this case may be as follows: QMFL
must differ from QM, but in such a way that in the low-energy limit
a density matrix in QMFL be coincident  with the density matrix in
QM. That is to say, QMFL is a deformation of QM and the parameter of
deformation depends on the measuring scale. This means that in QMFL
$\rho=\rho(x)$, where $x$ is the scale, and for $x\rightarrow\infty$
$\rho(x) \rightarrow \widehat{\rho}$, where $\widehat{\rho}$ is the
density matrix in QM.Since on the Planck's scale $Sp[\rho]<1$, then
for such scales $\rho=\rho(x)$, where $x$ is the scale, is not a
density matrix as it is generally defined in QM. On Planck's scale
$\rho(x)$ is referred to as  ``density pro-matrix''. As follows from
the above, the density matrix $\widehat{\rho}$ appears in the limit
\cite{shalyt1}-\cite{shalyt5}:
\begin{equation}\label{U12}
\lim\limits_{x\rightarrow\infty}\rho(x)\rightarrow\widehat{\rho},
\end{equation}
when  QMFL turns to QM.

Thus, on Planck's scale the density matrix is inadequate to obtain
all information about the mean values of operators. A ``deformed''
density matrix (or pro-matrix) $\rho(x)$ with $Sp[\rho]<1$ has to be
introduced because a missing part of information $1-Sp[\rho]$ is
encoded in the quantity $l^{2}_{min}/a^{2}$, whose specific weight
decreases as the scale $a$ expressed in  units of $l_{min}$ is going
up.

\section{QMFL as a deformation of QM}
\subsection{Main Definitions}
Here we describe QMFL as a deformation of QM using the
above-developed formalism of density pro-matrix. Within it density
pro-matrix should be understood as a deformed density matrix in
QMFL. As fundamental parameter of deformation we use the quantity
$\alpha=l_{min}^{2 }/x^{2}$, where $x$ is the scale. The following
deformation is not claimed as the only one satisfying all the
above properties. Of course, some other deformations are also
possible. At the same time, it seems most natural in a sense that
it allows for minimum modifications of the conventional density
matrix in QM.

\noindent {\bf Definition 1.}\cite{shalyt1}-\cite{shalyt5}

\noindent Any system in QMFL is described by a density pro-matrix
of the form $\rho(\alpha)=\sum_{i}\omega_{i}(\alpha)|i><i|$, where
\begin{enumerate}
\item $0<\alpha\leq1/4$.
\item Vectors $|i>$ form a full orthonormal system;
\item Coefficients $\omega_{i}(\alpha)\geq 0$ and for all $i$  the
 limit $\lim\limits_{\alpha\rightarrow
0}\omega_{i}(\alpha)=\omega_{i}$ exists;
\item
$Sp[\rho(\alpha)]=\sum_{i}\omega_{i}(\alpha)<1$,
$\sum_{i}\omega_{i}=1$.
\item For every operator $B$ and any $\alpha$ there is a
mean operator $B$ depending on  $\alpha$:
$$<B>_{\alpha}=\sum_{i}\omega_{i}(\alpha)<i|B|i>.$$
\end{enumerate}
Finally, in order that our definition 1 be in agreement with the
result of section 2, the following condition must be fulfilled:
\begin{equation}\label{U13}
Sp[\rho(\alpha)]-Sp^{2}[\rho(\alpha)]\approx\alpha.
\end{equation}
Hence we can find the value for $Sp[\rho(\alpha)]$ satisfying the
condition of definition 1:
\begin{equation}\label{U14}
Sp[\rho(\alpha)]\approx\frac{1}{2}+\sqrt{\frac{1}{4}-\alpha}.
\end{equation}

According to statement 5  $<1>_{\alpha}=Sp[\rho(\alpha)]$.
Therefore, for any scalar quantity $f$ we have $<f>_{\alpha}=f
Sp[\rho(\alpha)]$. In particular, the mean value
$<[x_{\mu},p_{\nu}]>_{\alpha}$ is equal to
\begin{equation}\label{U15sup}
<[x_{\mu},p_{\nu}]>_{\alpha}= i\hbar\delta_{\mu,\nu}
Sp[\rho(\alpha)].
\end{equation}
We denote the limit $\lim\limits_{\alpha\rightarrow
0}\rho(\alpha)=\rho$ as the density matrix. Evidently, in the
limit $\alpha\rightarrow 0$ we return to QM.

As follows from definition 1,
$<|j><j|>_{\alpha}=\omega_{j}(\alpha)$, from whence the
completeness condition by $\alpha$ is
\\$<(\sum_{i}|i><i|)>_{\alpha}=<1>_{\alpha}=Sp[\rho(\alpha)]$. The
norm of any vector $|\psi>$ assigned to  $\alpha$ can be defined
as

$$<\psi|\psi>_{\alpha}=<\psi|(\sum_{i}|i><i|)_{\alpha}|\psi>
=<\psi|(1)_{\alpha}|\psi>=<\psi|\psi> Sp[\rho(\alpha)],$$

 where
$<\psi|\psi>$ is the norm in QM, i.e. for $\alpha\rightarrow 0$.
Similarly, the described theory may be interpreted using a
probabilistic approach, however requiring  replacement of $\rho$
by $\rho(\alpha)$ in all formulae.

\renewcommand{\theenumi}{\Roman{enumi}}
\renewcommand{\labelenumi}{\theenumi.}
\renewcommand{\labelenumii}{\theenumii.}

\subsection{Some obvious implications}
It should be noted:

\begin{enumerate}
\item The above limit covers both Quantum
and Classical Mechanics. Indeed, since $\alpha\sim L_{p}^{2 }/x^{2
}=G \hbar/c^3 x^{2 }$, we obtain:
\begin{enumerate}
\item $(\hbar \neq 0,x\rightarrow
\infty)\Rightarrow(\alpha\rightarrow 0)$ for QM;
\item $(\hbar\rightarrow 0,x\rightarrow
\infty)\Rightarrow(\alpha\rightarrow 0)$ for Classical Mechanics;
\end{enumerate}
\item As a matter of fact, the deformation parameter $\alpha$
should assume the value $0<\alpha\leq1$.  However, as seen from
(\ref{U14}), $Sp[\rho(\alpha)]$ is well defined only for
$0<\alpha\leq1/4$.That is if $x=il_{min}$ and $i\geq 2$, then
there is not any problem. At the point where $x=l_{min}$ there is
a singularity related to complex values assumed by
$Sp[\rho(\alpha)]$ , i.e. to the impossibility of obtaining a
diagonalized density pro-matrix at this point over the field of
real numbers. For this reason definition 1 has no sense at the
point $x=l_{min}$.We will return to this question when considering
singularity and hypothesis of cosmic censorship in the following
section.
\item We consider possible solutions for (\ref{U13}).
For instance, one of the solutions of (\ref{U13}), at least to the
first order in $\alpha$, is $$\rho^{*}(\alpha)=\sum_{i}\alpha_{i}
exp(-\alpha)|i><i|,$$ where all $\alpha_{i}>0$ are independent of
$\alpha$  and their sum is equal to 1. In this way
$Sp[\rho^{*}(\alpha)]=exp(-\alpha)$. Indeed, we can easily verify
that \begin{equation}\label{U15}
Sp[\rho^{*}(\alpha)]-Sp^{2}[\rho^{*}(\alpha)]=\alpha+O(\alpha^{2}).
\end{equation}
The exponential ansatz for $\rho^{*}(\alpha)$ given here will be
included in subsequent sections. Note that in the momentum
representation $\alpha=p^{2}/p_{max}^{2}\sim p^{2}/p^{2}_{pl}$,
where $p_{pl}$ is the Planck's momentum. When present in matrix
elements, $exp(-\alpha)$ can damp the contribution of great
momenta in a perturbation theory.
\item It is clear that within the proposed description the
states with a unit probability, i.e. pure states, can appear only
in the limit $\alpha\rightarrow 0$, when all $\omega_{i}(\alpha)$
except one are equal to zero or when they tend to zero at this
limit. In our treatment pure states are states, which can be
represented in the form $|\psi><\psi|$, where $<\psi|\psi>=1$.

\item We suppose that all definitions concerning a
density matrix can be carried over to the above-mentioned
deformation of Quantum Mechanics (QMFL)  changing the density
matrix $\rho$ by the density pro-matrix $\rho(\alpha)$ with
subsequent passing to the low-energy limit $\alpha\rightarrow 0$.
Specifically, for statistical entropy we have
\begin{equation}\label{U16}
S_{\alpha}=-Sp[\rho(\alpha\ln(\rho(\alpha))].
\end{equation}
The quantity of $S_{\alpha}$ seems never to be equal to zero as
$\ln(\rho(\alpha))\neq 0$ and hence $S_{\alpha}$ may be equal to
zero at the limit $\alpha\rightarrow 0$ only.
\end{enumerate}
The following statements are essential for our study:
\begin{enumerate}
\item If we carry out a measurement at a pre-determined scale, it is
impossible to regard the density pro-matrix as a density matrix
with an accuracy better than the limit $\sim10^{-66+2n}$, where
$10^{-n}$ is the measuring scale. In the majority of known cases
this is sufficient to consider the density pro-matrix as a density
matrix. But at Planck's scale, where quantum gravitational effects
and Planck's energy levels cannot be neglected, the difference
between $\rho(\alpha)$ and  $\rho$ should be taken into
consideration.

\item Proceeding from the above, on Planck's scale the
notion of Wave Function of the Universe (as introduced in
\cite{r11}) has no sense, and quantum gravitation effects in this
case should be described with the help of density pro-matrix
$\rho(\alpha)$ only.
\item Since density pro-matrix $\rho(\alpha)$ depends on the measuring
scale, evolution of the Universe within the inflationary model
paradigm \cite{r10} is not a unitary process, or otherwise the
probabilities $p_{i}=\omega_{i}(\alpha)$  would be preserved.
\end{enumerate}

\section{Applications of the Quantum-Mechanical Density Pro-Matrix}
In this section some apparent applications of the primary
definitions and methods derived in the previous section are given
\cite{shalyt3}-\cite{shalyt5}.
\subsection{Dynamic aspects of QMFL. Deformed Liouville  equation}
 Let's suppose that in QMFL a
density pro-matrix has the form $\rho[\alpha(t),t]$, in other
words, it depends on two parameters: time $t$ and parameter of
deformation $\alpha$, which also depends on time
($\alpha=\alpha(t)$). Then, we have
\begin{equation}\label{U17}
\rho[\alpha(t),t]=\sum\omega_{i}[\alpha(t)]|i(t)><i(t)|.
\end{equation}
Differentiating the last expression with respect to time, we
obtain
\begin{equation}\label{U18}
\frac{d\rho}{dt}=\sum_{i}
\frac{d\omega_{i}[\alpha(t)]}{dt}|i(t)><i(t)|-i[H,\rho(\alpha)]=d[ln\omega(\alpha)]\rho
(\alpha)-i[H,\rho(\alpha)].
\end{equation}
Where $ln[\omega(\alpha)]$ is a row-matrix and $\rho(\alpha)$ is a
column-matrix. In such a way we have obtained a prototype of
 Liouville's equation.

Let's consider some  cases of particular importance.
\begin{enumerate}
\item First we consider the process of time
evolution at low energies, i.e. when $\alpha \approx 0$,
$\alpha(t)\approx 0$ and $t \to \infty$. Then it is clear that
$\omega_{i}(\alpha)\approx \omega_{i} \approx constant$. The first
term in (\ref{U18}) vanishes and we obtain  Liouville equation.
\item Also we obtain  the Liouville's equation when using
inflationary approach and going to large-scales. Within the
inflationary approach the scale $a \approx e^{Ht}$, where $H$ is
the Hubble's constant and $t$ is time. Therefore $\alpha \sim
e^{-2Ht}$ and when $t$ is quite big $\alpha\to 0$. In other words,
$\omega_{i}(\alpha) \to \omega_{i}$, the first term in (\ref{U18})
vanishes and again  we obtain the Liouville's equation.
\item At very early stage of the inflationary process  or even before it
took place $\omega_{i}(\alpha)$ was not a constant and hense, the
first term in (\ref{U18}) should be taking into account. This way
we obtain a deviation from the Liouville's equation.
\item Finally, let us consider the case when $\alpha(0) \approx 0$,
$\alpha(t)>0$ where $t \to \infty$. In this case we are going from
low-energy to high-energy scale one and $\alpha(t)$ grows when $t
\to \infty$. The first term in (\ref{U18}) is significant and we
obtain an addition to the Liouville's equation of the form
$$d[ln\omega(\alpha)]\rho(\alpha).$$ This could be the case  when
matter goes into a black hole and is moving in direction of the
singularity (to the Planck's scale).
\end{enumerate}

 \subsection{Singularity, entropy and information loss in black
holes}
Note that remark II in section 3.2 about complex meaning
assumed by the density pro-matrix at the point with fundamental
length is directly related to the singularity problem and cosmic
censorship in the General Theory of Relativity \cite{r12}.  For
instance, considering a Schwarzchild's black hole (\cite{r13})
with metrics:
\begin{equation}\label{U19}
 ds^2 = - (1 - \frac{2M}{r}) dt^2 +
\frac{dr^2}{(1 - \frac{2M}{r})} + r^2 d \Omega_{II}^2,
\end{equation}
we obtain a well-known a singularity at the point $r=0$. In our
approach this corresponds to the point with fundamental length
($r=l_{min}$). At this point we are not able to measure anything,
since at this point $\alpha=1$ and $Sp[\rho (\alpha)]$ becomes
complex. Thus, we carry out a measurement, starting from the point
$r=2l_{min}$ that corresponds to $\alpha=1/4$. Consequently, the
initial singularity $r=l_{min}$, which cannot be measured, is hidden
of observation. This could confirm the hypothesis of cosmic
censorship in this particular  case. By this hypothesis  "a naked
singularity cannot be observed". Thus, QMFL in our approach "feels"
the singularity  compared with QM, that does not
\cite{shalyt4,shalyt5}. Statistical entropy, associated with the
density pro-matrix and introduced in the remark V section 3 is
written $$S_{\alpha}=-Sp[\rho(\alpha)\ln(\rho(\alpha))],$$ and may
be interpreted as a density of entropy on the unit  minimal area
$l^{2}_{min}$ depending on the scale $x$. It could be quite big
close to the singularity, i.e. for $\alpha\rightarrow 1/4$. This
does not contradict the second law of Thermodynamics since maximal
entropy of a specific  object in the Universe is proportional to the
square of its surface $A$, measured in units of minimal square
$l^{2}_{min}$ or Planck's square $L_{p}^2$, as  shown in some papers
(see, for instance \cite{r14}). Therefore, in the expanded Universe
since surface $A$ grows, entropy does not decrease.

The obtained results enable one to consider anew the information
loss problem associated with black holes \cite{r15,r16}, at least,
for the case of "mini" black holes \cite{shalyt4,shalyt5}. Indeed,
when we consider these black holes, Planck's scale is a factor. It
was shown that entropy of matter absorbed by a black hole at this
scale is not equal to zero, supporting the data of R.Myers
\cite{Myers}. According to his results, the pure state cannot form a
black hole. Then,  it is necessary to reformulate the problem per
se, since so far in all publications on information paradox zero
entropy at the initial state has been compared to  nonzero entropy
at the final state. According to our analysis at the Planck's scale
there is not initial zero entropy and "mini" black holes with masses
of the order $M_{pl}$ should not radiate at all. Similar results
were deduced by A.D.Helfer\cite{r31} using another approach:
"p.1...The possibility that non-radiating "mini" black holes should
be taken seriously; such holes could be part of the dark matter in
the Universe". Note that in \cite{r31} the main argument in favor of
the existence of non-radiating "mini" black holes developed with
consideration of quantum gravity effects. In our analysis these
effects are considered implicitly since,as stated above, any
approach in quantum gravity leads to the fundamental-length concept
\cite{r1}. Besides, it should be noted that in some recent papers
for all types of black holes QMFL with GUR is considered from the
start \cite{r18},\cite{r32}. By this approach stable remnants with
masses of the order of Planck's ones $M_{pl}$ emerge during the
process of black hole evaporation. From here it follows that black
holes should not evaporate fully. We arrive to the conclusion that
results given in \cite{r13,r19} are correct only in the
semi-classical approximation and they should not be applicable to
the quantum back hole analysis.

At least at a qualitative level the above results can clear up the
answer to the question, how information may be lost at big black
holes formed due to the star collapse. Our point of view is close to
that of R.Penrose's one \cite{r20} who considers  that information
in black holes is lost when matter meets a singularity. In our
approach information loss takes place in the same form. Indeed, near
the horizon of events an approximately pure state with the initial
entropy practically equal to zero $S^{in}=-Sp[\rho\ln(\rho)]$, that
corresponds to $\alpha \to 0$, when approaching a singularity (of
reaching the Planck's scale) gives yet non zero entropy
$S_{\alpha}=-Sp[\rho(\alpha)\ln(\rho(\alpha))]>0$ for $\alpha
>0$. Therefore, entropy increases and information is lost in this
black hole. We can (at the moment, at a qualitative level)
evaluate the entropy of black holes. Actually, starting from a
density matrix for the pure state at the "entry" to a black hole
$\rho_{in}=\rho_{pure}$ with zero entropy $S^{in}=0$, we obtain
with a straightforward "naive" calculation (that is (\ref{U13}) is
considered  an exact relation). Then,for the singularity in the
black hole the corresponding entropy of the density pro-matrix
$Sp[\rho_{out}]=1/2$ at $\alpha=1/4$ is $$S^{out}=S_{1/4}=-1/2
\ln1/2 \approx 0.34657.$$
 Taking into account that total entropy of a
black hole is proportional to the quantum area of surface A,
measured in Planck's units of area $L_{p}^2$ \cite{r21}, we obtain
the following value for quantum entropy of a black hole:
\begin{equation}\label{U20}
S_{BH}= 0.34657 \frac{A}{L_{p}^2}
\end{equation}
This formula differs from the well-known one given by
Bekenstein-Hawking for black hole entropy $S_{BH}=\frac{1}{4}
\frac{A}{L_{p}^2}$ \cite{r22}. This result was obtained in the
semi-classical approximation. At the present moment quantum
corrections to this formula are an object of investigation
\cite{r23}. As it was mentioned above we carry out straightforward
calculation. Otherwise, using the ansatz of statement  remark III
in section 3 and assuming that spur of density pro-matrix is equal
to $Sp[\rho^{*}(\alpha)]=exp(-\alpha)$, we obtain for $\alpha=1/4$
that entropy is equal to $$S^{* out}=S^{*}_{1/4}=-Sp[exp(-1/4)\ln
exp(-1/4)]\approx 0.1947,$$ and consequently we arrive to the
following value of entropy
\begin{equation}\label{U21}
S_{BH} = 0.1947 \frac{A}{L_{p}^2}
\end{equation}
that is the closest  the result obtained in \cite{r23}. Our
approach leading to formula (\ref{U21}) is from the very beginning
based on quantum nature of black holes. Note here that in the
approaches used up to now to modify Liouville's equation due to
information paradox \cite{r24} the added member appearing in the
right side of (\ref{U18}) takes the form $$-\frac{1}{2}\sum_{\xi
\gamma \neq 0} (Q^{\gamma}Q^{\xi}\rho+\rho Q^{\gamma}Q^{\xi}-2
Q^{\xi}\rho Q^{\gamma}),$$ where $Q^{\xi}$
  is a full orthogonal set
of Hermitian matrices with $Q^{0} =1$. In this case either locality
or conservation of energy-impulse tensor is broken down. By the
approach offered in this paper the member added in the deformed
Liouville's equation,in our opinion, has a more natural and
beautiful form: $$d[ln\omega(\alpha)]\rho (\alpha).$$ In the limit
$\alpha\to 0$ all properties of QM are conserved, the added member
vanishes and we obtain Liouville's equation.

The information paradox problem at black holes is considered in
greater detail in section 7, where the above methods provide a new
approach to  this problem.

\subsection{Bekenstein-Hawking formula}
The problem is whether can the well-known semiclassical
Bekenstein-Hawking formula for Black Hole  entropy
\cite{r21},\cite{r32} can be obtained within the proposed approach
? We show how  to do it \cite{shalyt5}. To obtain black hole
quantum entropy, we use the formula
  $S_{\alpha}=-Sp[\rho(\alpha)\ln(\rho(\alpha))]=-<\ln(\rho(\alpha))>_{\alpha}$
    when $\alpha$ takes its maximal meaning ($\alpha = 1/4$).
   In this case (\ref{U20}) and (\ref{U21}) can be written as
\begin{equation}\label{U22}
S_{BH} = -<\ln(\rho(1/4))>_{1/4} \frac{A}{L_{p}^2},
\end{equation}
for different $\rho(\alpha)$ in  (\ref{U20}) and  (\ref{U21}) but
for the same value of $\alpha$ ($\alpha = 1/4$). Semiclassical
approximation works only at large-scales, therefore measuring
procedure is also defined at large scales. In other words, all mean
values must be taken when $\alpha = 0$. However, for the operators
whose mean values are calculated the dependence on $\alpha$ should
be taken  into account since according to the well-known Hawking's
paper \cite{r14}, operator of superscattering $\$$ translates
$\$:\rho_{in}\mapsto\rho_{out}$, where in the case considered
$\rho_{in}=\rho_{pure}$ and $\rho_{out}=\rho_{pure}^{*}(\alpha)=
exp(-\alpha)\rho_{pure}=exp(-1/4)\rho_{pure}$ conforming to the
exponential ansatz of statement  III, section 3. Therefore we have
$$S^{semiclass}_{\alpha}=-<\ln(\rho(\alpha))>$$
and formula for semiclassical entropy of a black hole takes the
form
\begin{equation}\label{U23}
S^{semiclass}_{BH} = -<\ln(\rho(1/4))>
\frac{A}{L_{p}^2}=-<ln[exp(-1/4)]\rho_{pure}>\frac{A}{L_{p}^2}
=\frac{1}{4}\frac{A}{L_{p}^2}
\end{equation}
that coincides with the well-known Bekenstein-Hawking formula. It
should be noted that  $\alpha = 1/4$  in our approach appears in
section 3 quite naturally as a maximal meaning for which
$Sp\rho(\alpha)$ still stays real, according to (\ref{U13}) and
(\ref{U14}). Apparently, if considering corrections of order
higher than 1 on $\alpha$, then members from $O(\alpha^{2})$ in
the formula for $\rho_{out}$ in (\ref{U15}) can give quantum
corrections \cite{r23} for $S^{semiclass}_{BH}$ (\ref{U23}) in our
approach.

\subsection{Some comments on Shr{\"o}dinger's picture}
 As it was indicated above in the
statement 1 section 3.2, we are able to obtain from QMFL two
limits: Quantum and Classical Mechanics. The deformation described
here should be understood as "minimal" in the sense that we have
deformed only the probability $\omega_{i}\rightarrow
\omega_{i}(\alpha)$, whereas the state vectors have been not
deformed. In a most complete treatment we have to consider vectors
$|i(\alpha)><i(\alpha)|$ instead $|i><i|$,  and in this case the
full picture will be very complicated. It is easy to understand
how Shrodinger's picture is transformed in QMFL \cite{shalyt5}.
The prototype of Quantum Mechanical normed wave function $\psi(q)$
with $\int|\psi(q)|^{2}dq=1$ in QMFL is $\theta(\alpha)\psi(q)$.
The deformation parameter   $\alpha$ assumes the value
$0<\alpha\leq1/4$. Its properties are
$|\theta(\alpha)|^{2}<1$,$\lim\limits_{\alpha\rightarrow
0}|\theta(\alpha)|^{2}=1$ and the relation
$|\theta(\alpha)|^{2}-|\theta(\alpha)|^{4}\approx \alpha$ takes
place. In such a way the full probability always is less than 1:
$p(\alpha)=|\theta(\alpha)|^{2}=\int|\theta(\alpha|^{2}|\psi(q)|^{2}dq<1$
tending to 1 when  $\alpha\rightarrow 0$. In the most general case
of arbitrarily normed state in QMFL
$\psi=\psi(\alpha,q)=\sum_{n}a_{n}\theta_{n}(\alpha)\psi_{n}(q)$
with $\sum_{n}|a_{n}|^{2}=1$ the full probability is
$p(\alpha)=\sum_{n}|a_{n}|^{2}|\theta_{n}(\alpha)|^{2}<1$ and
 $\lim\limits_{\alpha\rightarrow 0}p(\alpha)=1$.

 It is natural that in QMFL Shrodinger's equation is also
deformed. It is replaced by  equation
\begin{equation}\label{U24}
\frac{\partial\psi(\alpha,q)}{\partial t}
=\frac{\partial[\theta(\alpha)\psi(q)]}{\partial
t}=\frac{\partial\theta(\alpha)}{\partial
t}\psi(q)+\theta(\alpha)\frac{\partial\psi(q)}{\partial t},
\end{equation}
where the second term in the right side generates the Shrodinger's
equation since
\begin{equation}\label{U25}
\theta(\alpha)\frac{\partial\psi(q)}{\partial
t}=\frac{-i\theta(\alpha)}{\hbar}H\psi(q).
\end{equation}

Here $H$ is the Hamiltonian and the first member is added,
similarly to the member appearing in the deformed Loiuville's
equation and  vanishing when $\theta[\alpha(t)]\approx const$. In
particular, this takes place in the low energy limit in QM, when
$\alpha\rightarrow 0$.  Note that the above theory  is not a
time-reversal as QM, since the combination $\theta(\alpha)\psi(q)$
breaks down this property in the deformed Shrodinger's equation.
Time-reversal is conserved only in the low energy limit, when
quantum mechanical Shrodinger's equation is valid.

\section{Density Matrix Deformation in Statistical Mechanics of
Early Universe}
\subsection{Main definition and properties}
First we revert to  the Generalized Uncertainty Relations
 "coordinate - momentum" (section 2,formula (\ref{U2})) :
\begin{equation}\label{U5s}
\triangle x\geq\frac{\hbar}{\triangle p}+\alpha^{\prime}
L_{p}^2\frac{\triangle p}{\hbar}.
\end{equation}
 Using relations (\ref{U5s}) it is easy to obtain a similar relation for the
 "energy - time" pair. Indeed (\ref{U5s}) gives
\begin{equation}\label{U6s}
\frac{\Delta x}{c}\geq\frac{\hbar}{\Delta p c }+\alpha^{\prime}
L_{p}^2\,\frac{\Delta p}{c \hbar},
\end{equation}
then
\begin{equation}\label{U7s}
\Delta t\geq\frac{\hbar}{\Delta
E}+\alpha^{\prime}\frac{L_{p}^2}{c^2}\,\frac{\Delta p
c}{\hbar}=\frac{\hbar}{\Delta E}+\alpha^{\prime}
t_{p}^2\,\frac{\Delta E}{ \hbar}.
\end{equation}
where the smallness of $L_p$ is taken into account so that the
difference between $\Delta E$ and $\Delta (pc)$ can be neglected and
$t_{p}$  is the Planck time $t_{p}=L_p/c=\sqrt{G\hbar/c^5}\simeq
0,54\;10^{-43}sec$. From whence it follows that we have a  maximum
energy of the order of Planck's:
$$E_{max}\sim E_{p}$$
Proceeding to the Statistical Mechanics, we further assume that an
internal energy of any ensemble U could not be in excess of
$E_{max}$ and hence temperature $T$ could not be in excess of
$T_{max}=E_{max}/k_{B}\sim T_{p}$. Let us consider density matrix in
Statistical Mechanics (see \cite{r34}, Section 2, Paragraph 3):
\begin{equation}\label{U8s}
\rho_{stat}=\sum_{n}\omega_{n}|\varphi_{n}><\varphi_{n}|,
\end{equation}
where the probabilities are given by
$$\omega_{n}=\frac{1}{Q}\exp(-\beta E_{n})$$ and
$$Q=\sum_{n}\exp(-\beta E_{n}).$$
Then for a canonical Gibbs ensemble the value
\begin{equation}\label{U9s}
\overline{\Delta(1/T)^{2}}=Sp[\rho_{stat}(\frac{1}{T})^{2}]
-Sp^{2}[\rho_{stat}(\frac{1}{T})],
\end{equation}
is always equal to zero, and this follows from the fact that
$Sp[\rho_{stat}]=1$. However, for very high temperatures $T\gg0$
we have $\Delta (1/T)^{2}\approx 1/T^{2}\geq 1/T_{max}^{2}$. Thus,
for $T\gg0$ a statistical density matrix $\rho_{stat}$ should be
deformed so that in the general case \cite{shalyt6,shalyt7}
\begin{equation}\label{U10s}
Sp[\rho_{stat}(\frac{1}{T})^{2}]-Sp^{2}[\rho_{stat}(\frac{1}{T})]
\approx \frac{1}{T_{max}^{2}},
\end{equation}
or \begin{equation}\label{U11s}
Sp[\rho_{stat}]-Sp^{2}[\rho_{stat}] \approx
\frac{T^{2}}{T_{max}^{2}}.
\end{equation}
In this way $\rho_{stat}$ at very high $T\gg 0$ becomes dependent on
the parameter $\tau = T^{2}/T_{max}^{2}$, i.e. in the most general
case
$$\rho_{stat}=\rho_{stat}(\tau)$$ and $$Sp[\rho_{stat}(\tau)]<1$$
and for $\tau\ll 1$ we have $\rho_{stat}(\tau)\approx\rho_{stat}$
(formula (\ref{U8s})) .\\ This situation is identical to the case
associated with the deformation parameter $\alpha=
l_{min}^{2}/x^{2}$ of QMFL given in section 3. That is the
condition $Sp[\rho_{stat}(\tau)]<1$ has an apparent physical
meaning when:
\begin{enumerate}
 \item At temperatures close to $T_{max}$ some portion of information
about the ensemble is inaccessible in accordance with the
probability that is less than unity, i.e. incomplete probability.
 \item And vice versa, the longer is the distance from $T_{max}$ (i.e.
when approximating the usual temperatures), the greater is the
bulk of information and the closer is the complete probability to
unity.
\end{enumerate}
 Therefore similar to the introduction of the deformed
quantum-mechanics density matrix in section 3 we give the
following

\noindent {\bf Definition 2.} {\bf(Deformation of Statistical
Mechanics)}\cite{shalyt6,shalyt7,shalyt8}

\noindent Deformation of Gibbs distribution valid for temperatures
on the order of the Planck's $T_{p}$ is described
 by deformation of a statistical density matrix
  (statistical density pro-matrix) of the form
$${\bf \rho_{stat}(\tau)=\sum_{n}\omega_{n}(\tau)|\varphi_{n}><\varphi_{n}|}$$
 having the deformation parameter
$\tau = T^{2}/T_{max}^{2}$, where
\begin{enumerate}
\item $0<\tau \leq 1/4$.
\item The vectors $|\varphi_{n}>$ form a full orthonormal system;
\item $\omega_{n}(\tau)\geq 0$ and for all $n$ at $\tau \ll 1$
 we obtain
 $\omega_{n}(\tau)\approx \omega_{n}=\frac{1}{Q}\exp(-\beta E_{n})$
In particular, $\lim\limits_{T_{max}\rightarrow \infty
(\tau\rightarrow 0)}\omega_{n}(\tau)=\omega_{n}$
\item
$Sp[\rho_{stat}]=\sum_{n}\omega_{n}(\tau)<1$,
$\sum_{n}\omega_{n}=1$;
\item For every operator $B$ and any $\tau$ there is a
mean operator $B$ depending on  $\tau$
$$<B>_{\tau}=\sum_{n}\omega_{n}(\tau)<n|B|n>.$$
\end{enumerate}
Finally, in order that our Definition 2 agree with the formula
(\ref{U11s}), the following condition must be fulfilled:
\begin{equation}\label{U12s}
Sp[\rho_{stat}(\tau)]-Sp^{2}[\rho_{stat}(\tau)]\approx \tau.
\end{equation}
Hence we can find the value for $Sp[\rho_{stat}(\tau)]$
 satisfying
the condition of Definition 2 (similar to Definition 1):
\begin{equation}\label{U13s}
Sp[\rho_{stat}(\tau)]\approx\frac{1}{2}+\sqrt{\frac{1}{4}-\tau}.
\end{equation}
It should be noted:

\begin{enumerate}
\item The condition $\tau \ll 1$ means that $T\ll T_{max}$ either
$T_{max}=\infty$ or both in accordance with a normal Statistical
Mechanics and canonical Gibbs distribution (\ref{U8s})
\item Similar to QMFL in Definition 1, where the deformation
parameter $\alpha$ should assume the value $0<\alpha\leq1/4$. As
seen from (\ref{U13s}), here $Sp[\rho_{stat}(\tau)]$ is well
defined only for $0<\tau\leq1/4$. This means that the feature
occurring in QMFL at the point of the fundamental length
$x=l_{min}$ in the case under consideration is associated with the
fact that {\bf highest  measurable temperature of the ensemble is
always} ${\bf T\leq \frac{1}{2}T_{max}}$.

\item The constructed deformation contains all four fundamental constants:
 $G,\hbar,c,k_{B}$ as $T_{max}=\varsigma T_{p}$,where $\varsigma$
 is the denumerable function of  $\alpha^{\prime}$
(\ref{U5s})and $T_{p}$, in its turn, contains all the
above-mentioned
 constants.

\item Again similar to QMFL, as a possible solution for (\ref{U12s})
we have an exponential ansatz
$$\rho_{stat}^{*}(\tau)=\sum_{n}\omega_{n}(\tau)|n><n|=\sum_{n}
exp(-\tau) \omega_{n}|n><n|$$
\begin{equation}\label{U14s}
Sp[\rho_{stat}^{*}(\tau)]-Sp^{2}[\rho_{stat}^{*}(\tau)]=\tau+O(\tau^{2}).
\end{equation}
In such a way with the use of an exponential ansatz (\ref{U14s})
the deformation of a canonical Gibbs distribution at Planck scale
(up to factor $1/Q$) takes an elegant and completed form:
\begin{equation}\label{U15s}
{\bf \omega_{n}(\tau)=exp(-\tau)\omega_{n}= exp(-\frac{T^{2}}
{T_{max}^{2}}-\beta E_{n})}
\end{equation}
where $T_{max}= \varsigma T_{p}$
\end{enumerate}
\subsection{Some implications}
Using in this section only the
exponential ansatz of (\ref{U14s}), in the coordinate
representation we have the following:
\begin{equation}\label{U16s}
\rho(x,x^{\prime},\tau)=\sum_{i}\frac{1}{Q}e^{-\beta
E_{i}-\tau}\varphi_{i}(x)\varphi_{i}^{*}(x^{\prime})
\end{equation}
However, as $H \mid \varphi_{i}>=E_{i} \mid \varphi_{i}>$, then
\begin{equation}\label{U17s}
\rho(\beta,\tau)=\frac{1}{Q}\sum_{i}e^{-\beta H-\tau}\mid
\varphi_{i}><\varphi_{i}\mid=\frac{e^{-\beta H-\tau}}{Q},
\end{equation}
where $Q=\sum_{i}e^{-\beta E_{i}}=Spe^{-\beta H}$. Consequently,
\begin{equation}\label{U18s}
\rho(\beta,\tau)=\frac{e^{-\beta H-\tau}}{Spe^{-\beta H}}
\end{equation}
In this way the "deformed" average energy of a system is obtained
as
\begin{equation}\label{U19s}
U_{\tau}=Sp\rho(\tau)H=\frac{He^{-\beta H-\tau}}{Spe^{-\beta H}}
\end{equation}
The calculation of "deformed" entropy is also a simple task.
Indeed, in the general case of the canonical Gibbs distribution
the probabilities are equal to
\begin{equation}\label{U20s}
P_{n}=\frac{1}{Q}e^{-\beta E_{n}}
\end{equation}
Nevertheless, in case under consideration they are "replenished"
by $exp(-\tau)$ factor and hence are equal to
\begin{equation}\label{U21s}
P^{\tau}_{n}=\frac{1}{Q}e^{-(\tau+\beta E_{n})}.
\end{equation}
Thus, a new formula for entropy in this case is as follows:
\begin{equation}\label{U22s}
S_{\tau}=-k_{B}e^{-\tau}\sum_{n}P_{n}(lnP_{n}-\tau)
\end{equation}
It is obvious that
 $\lim\limits_{\tau\rightarrow 0}S_{\tau}=
S$, where $S$ - entropy of the canonical ensemble, that is a
complete analog of its counterpart in quantum mechanics at the
Planck scale $\lim\limits_{\alpha\rightarrow 0}S_{\alpha}= S$, where
$S$ - statistical entropy in quantum mechanics, and deformation
parameter $\tau$ is changed by $\alpha$ of section 3.

Given the average energy deformation in a system $U_{\tau}$ and
knowing the entropy deformation, one is enabled to calculate the
"deformed" free energy $F_{\tau}$ as well:
\begin{equation}\label{U23s}
F_{\tau}=U_{\tau}-TS_{\tau}
\end{equation}
Consider the counterpart of Liouville equation \cite{r34} for the
unnormed $\rho(\beta,\tau)$ (\ref{U18s}):
\begin{equation}\label{U24s}
-\frac{\partial\rho(\beta,\tau)}{\partial\beta}=
-\frac{\partial}{\partial\beta}e^{-\tau-\beta H},
\end{equation}
where
$$\tau=\frac{T^{2}}{T_{max}^{2}}=\frac{\beta^{2}_{max}}{\beta^{2}},$$
where $\beta_{max}=1/k_{B}T_{max}\sim 1/k_{B}T_{P}\equiv
\beta_{P}$, $\tau=\tau(\beta)$. Taking this into consideration and
expanding the right-hand side of equation (\ref{U24s}), we get
deformation  of Liouville equation further referred to as
$\tau$-deformation:
\begin{equation}\label{U25s}
-\frac{\partial\rho(\beta,\tau)}{\partial\beta}=
-e^{-\tau}\frac{\partial \tau}{\partial \beta}+e^{-\tau}H
\rho(\beta)=e^{-\tau}[H \rho(\beta)-\frac{\partial \tau}{\partial
\beta}],
\end{equation}
where $\rho(\beta)=\rho(\beta,\tau=0)$.
 The first term in brackets (\ref{U25s}) generates Liouville equation.
Actually, taking the limit of the left and right sides
(\ref{U25s}) for $\tau\rightarrow 0$, we derive the normal
Liouville equation for $\rho(\beta)$ in statistical mechanics
\cite{r34}:
\begin{equation}\label{U26s}
-\frac{\partial\rho(\beta)}{\partial\beta}=H \rho(\beta)
\end{equation}
By this means we obtain a complete analog of the
quantum-mechanical results for the associated deformation of
Liouville equation derived  in section 4.1 and
\cite{shalyt3}-\cite{shalyt5}.
 Namely:
\\
(1)Early Universe (scales approximating those of the Planck's,
original singularity, $\tau>0$). The density pro-matrix
$\rho(\beta,\tau)$ is introduced and a $\tau$-deformed Liouville
equation(\ref{U25s}), respectively;
\\
(2)after the inflation extension (well-known scales, $\tau\approx
0$) the normal density matrix $\rho(\beta)$ appears in the limit
$\lim\limits_{\tau\rightarrow 0}\rho(\beta,\tau)=\rho(\beta)$.
$\tau$- deformation of Liouville equation (\ref{U25s}) is changed
by a well-known Liouville equation(\ref{U26s});
\\
(3)and finally the case of the matter absorbed by a black hole and
its tendency to the singularity. Close to the black hole
singularity both quantum and statistical mechanics are subjected
to deformation as they do in case of the original singularity
\cite{shalyt3}-\cite{shalyt5}. Introduction of temperature on the
order of the Planck's \cite{Castro1},\cite{Castro2} and hence the
deformation parameter $\tau > 0$ may be taken as an indirect
evidence for the fact. Because of this, the case is associated
with the reverse transition from the well-known density matrix in
statistical mechanics $\rho(\beta)$ to its $\tau$-deformation
$\rho(\beta,\tau)$ and from Liouville equation(\ref{U26s}) to its
$\tau$-deformation (\ref{U25s}).

\section{Generalized Uncertainty Relation
in Thermodynamics}

 Now we consider the thermodynamic uncertainty relations between the
inverse temperature and interior energy of a macroscopic ensemble
\begin{equation}\label{U12t}
\Delta \frac{1}{T}\geq\frac{k}{\Delta U},
\end{equation}
where $k$ is the Boltzmann constant.

 N.Bohr \cite{Bohr1} and
W.Heisenberg \cite{Heis1} first pointed out that such kind of
uncertainty principle should take place in thermodynamics. The
thermodynamic uncertainty  relations (\ref{U12t})  were proved by
many authors and in various ways \cite{Uncert1}. Therefore their
validity does not raise any doubts. Nevertheless, relation
(\ref{U12t}) was established using a standard model for the
infinite-capacity heat bath encompassing the ensemble. But it is
obvious from the above inequalities that at very high energies the
capacity of the heat bath can no longer be assumed infinite at the
Planck scale. Indeed, the total energy of the pair heat bath -
ensemble may be arbitrary large but finite, merely as the Universe
is born at a finite energy. Thus the quantity that can be
interpreted as a temperature of the ensemble must have the upper
limit and so does its main quadratic deviation. In other words the
quantity $\Delta (1/T)$ must be bounded from below. But in this case
an additional term should be introduced into
(\ref{U12t})\cite{shalyt9, shalyt10, shalyt7}
\begin{equation}\label{U12at}
\Delta \frac{1}{T}\geq\frac{k}{\Delta U} + \eta\,\Delta U,
\end{equation}
where $\eta$ is a coefficient. Dimension and symmetry reasons give
$$ \eta \sim \frac{k}{E_p^2}\enskip or\enskip \eta =
\alpha^{\prime} \frac{k}{E_p^2} $$
 As in the previous cases
inequality (\ref{U12at}) leads to the fundamental (inverse)
temperature.
\begin{equation}\label{U15t}
T_{max}=\frac{\hbar}{2\surd \alpha^{\prime}t_{p}
k}=\frac{\hbar}{\Delta t_{min} k}, \quad \beta_{min} = {1\over
kT_{max}} =  \frac{\Delta t_{min}}{\hbar}
\end{equation}
It should be noted that the same conclusion about the existence of
 maximal temperature in Nature can be made also considering black
hole evaporation \cite{Castro3}.

  Thus, we obtain the
system of generalized uncertainty relations in the symmetric form
\begin{equation}\label{U17t}
\left\{
\begin{array}{lll}
\Delta x & \geq & \frac{\displaystyle\hbar}{\displaystyle\Delta
p}+ \alpha^{\prime} \left(\frac{\displaystyle\Delta
p}{\displaystyle P_{pl}}\right)\,
\frac{\displaystyle\hbar}{\displaystyle P_{pl}}+... \\ &  &  \\
\Delta t & \geq & \frac{\displaystyle\hbar}{\displaystyle\Delta
E}+\alpha^{\prime} \left(\frac{\displaystyle\Delta
E}{\displaystyle E_{p}}\right)\,
\frac{\displaystyle\hbar}{\displaystyle E_{p}}+...\\
  &  &  \\
  \Delta \frac{\displaystyle 1}{\displaystyle T}& \geq &
  \frac{\displaystyle k}{\displaystyle\Delta U}+\alpha^{\prime}
  \left(\frac{\displaystyle\Delta U}{\displaystyle E_{p}}\right)\,
  \frac{\displaystyle k}{\displaystyle E_{p}}+...
\end{array} \right.
\end{equation}
or in the equivalent form
\begin{equation}\label{U18t}
\left\{
\begin{array}{lll}
\Delta x & \geq & \frac{\displaystyle\hbar}{\displaystyle\Delta
p}+\alpha^{\prime} L_{p}^2\,\frac{\displaystyle\Delta
p}{\displaystyle\hbar}+... \\
  &  &  \\
  \Delta t & \geq &  \frac{\displaystyle\hbar}{\displaystyle\Delta E}+\alpha^{\prime}
  t_{p}^2\,\frac{\displaystyle\Delta E}{\displaystyle\hbar}+... \\
  &  &  \\

  \Delta \frac{\displaystyle 1}{\displaystyle T} & \geq &
  \frac{\displaystyle k}{\displaystyle\Delta U}+\alpha^{\prime}
  \frac{\displaystyle 1}{\displaystyle T_{p}^2}\,
  \frac{\displaystyle\Delta U}{\displaystyle k}+...
\end{array} \right.
\end{equation}
where dots mean the existence of higher order corrections as in
\cite{r27}.
 Here $T_{p}$ is the Planck temperature:
$T_{p}=\frac{E_{p}}{k}$.

In conclusion of this section we would like to note that the
restriction on the heat bath made above makes the equilibrium
partition function non-Gibbsian \cite{r35}.

 Note that the last-mentioned inequality is symmetrical to the second one
with respect to substitution \cite{Castro4}
$$ t\mapsto\frac{1}{T}, \hbar\mapsto k,\triangle E\mapsto
\triangle U . $$ However this observation can by no means be
regarded as a rigorous proof of the generalized uncertainty relation
in thermodynamics.

There is  reason to believe that  rigorous justification for the
latter  (thermodynamic) inequalities in systems (\ref{U17t}) and
(\ref{U18t}) may be made by means of a certain deformation of Gibbs
distribution. One of such deformations that, by the author's
opinion, is liable to give the indicated result has been considered
in the previous section  of this chapter and in some other papers
\cite{shalyt6,shalyt7}.

\section{Non-Unitary and Unitary Transitions
 in Generalized Quantum  Mechanics
 and Information Problem Solving}
In this section the earlier obtained results are used for the
unitarity study in Generalized Quantum  Mechanics and
 Information Paradox Problem \cite{r15},\cite{r13},\cite{r16}.
 It is demonstrated that the existence of
black holes in the suggested approach in the end twice causes
nonunitary transitions resulting in the unitarity. In parallel
this problem is considered in other terms: entropy density,
Heisenberg algebra deformation terms, respective deformations of
Statistical Mechanics, - all showing the identity of the basic
results. From this an explicit solution for Information Paradox
Problem has been derived. This section is based on the results
presented in \cite{shalyt11,shalyt12,shalyt13}

\subsection{Some comments and unitarity in
QMFL}
 As has been indicated in section 4.4, time reversal
is retained in the large-scale limit only. The same is true for the
superposition principle in Quantum Mechanics. Indeed, it may be
retained in a very narrow interval of cases for the functions
$\psi_{1}(\alpha,q)=\theta(\alpha)\psi_{1}(q)$
$\psi_{2}(\alpha,q)=\theta(\alpha)\psi_{2}(q)$ with the same value
$\theta(\alpha)$. However, as for all $\theta(\alpha)$, their limit
is $\lim\limits_{\alpha\rightarrow 0}|\theta(\alpha)|^{2}=1$ or
equivalently $\lim\limits_{\alpha\rightarrow 0}|\theta(\alpha)|=1$,
in going to the low-energy limit each wave function $\psi(q)$ is
simply multiplied by the phase factor $\theta(0)$. As a result we
have Hilbert Space wave functions in QM. Comparison of both pictures
(Neumann's and Shr{\"o}dinger's) is indicative of the fact that
unitarity means the retention of the probabilities
$\omega_{i}(\alpha)$ or retention of the squared modulus (and hence
the modulus) for the function $\theta(\alpha)$:
$|\theta(\alpha)|^{2}$,($|\theta(\alpha)|$).That is
$$\frac{d\omega_{i}[\alpha(t)]}{dt}=0$$ or
$$\frac{d|\theta[\alpha(t)]|}{dt}=0.$$
In this way a set of unitary transformations of QMFL includes a
group $U$ of the unitary transformations for the wave functions
$\psi(q)$ in QM.

It is seen that on going from Planck's scale to the conventional one
, i.e. on transition from the Early Universe to the current one, the
scale has been rapidly changing in the process of inflation
expansion and the above conditions failed to be fulfilled:
\begin{equation}\label{U26h}
\frac{d\omega_{i}[\alpha(t)]}{dt}\neq 0, {\sloppy}
\frac{d|\theta[\alpha(t)]|}{dt}\neq 0.
\end{equation}
In terms of the density pro-matrices of sections 2,3 this is a
limiting transition from the density pro-matrix in QMFL
$\rho(\alpha)$,$\alpha>0$ , that is a prototype of the pure state at
$\alpha\rightarrow 0$ to the density matrix $\rho(0)=\rho$
representing a pure state in QM. Mathematically this means that a
nontotal probability (below 1) is changed by the total one (equal to
1). For the wave functions in Schr{\"o}dinger picture this limiting
transition from QMFL to QM is as follows:
$$\lim\limits_{\alpha\rightarrow 0}\theta(\alpha)\psi(q)=\psi(q)$$
up to the phase factor.

It is apparent that the above transition from QMFL to QM is not a
unitary process, as indicated in \cite{shalyt1}-\cite{shalyt5} and
section 3.2. However, the unitarity may be recovered when we
consider in a sense a reverse process: absorption of the matter by a
black hole and its transition to singularity conforming to the
reverse and nonunitary transition from QM to QMFL. Thus, nonunitary
transitions occur in this picture twice:
$$I.(QMFL,OS,\alpha\approx 1/4)\stackrel{Big\enskip
Bang}{\longrightarrow}(QM,\alpha\approx 0)$$
$$II.(QM,\alpha\approx 0)\stackrel{absorbing\enskip BH
}{\longrightarrow}(QMFL,SBH,\alpha\approx 1/4).$$ Here the following
abbreviations are used: OS for the Origin Singularity; BH for a
Black Hole; SBH for the Singularity in Black Hole.

 As a result of
these two nonunitary transitions, the total unitarity may be
recovered:
\\
\\
$$III.(QMFL,OS,\alpha\approx
1/4){\longrightarrow}(QMFL,SBH,\alpha\approx 1/4).$$
\\
\\
In such a manner the total information quantity in the Universe
remains unchanged, i.e. no information loss occurs.

 In terms of the deformed Liouville equation \cite{shalyt3}-\cite{shalyt5}
and section 4.1 we arrive to the expression with the same
right-hand parts for $t_{initial}\sim t_{Planck}$ and $t_{final}$
(for $\alpha\approx 1/4$).
\begin{eqnarray}\label{U27h}
\frac{d\rho[\alpha(t),t]}{dt}=\sum_{i}
\frac{d\omega_{i}[\alpha(t)]}{dt}|i(t)><i(t)|-\nonumber \\
-i[H,\rho(\alpha)]= d[ln\omega(\alpha)]\rho
(\alpha)-i[H,\rho(\alpha)].
\end{eqnarray}
It should be noted that for the closed Universe one can consider
Final Singularity (FS) rather than the Singularity of Black Hole
(SBH), and then the right-hand parts of diagrams II and III will be
changed:
$$IIa.(QM,\alpha\approx 0)\quad\underrightarrow{\text{Big
Crunch}}\quad (QMFL,FS,\alpha\approx 1/4),$$
$$IIIa.(QMFL,OS,\alpha\approx
1/4){\longrightarrow}(QMFL,FS,\alpha\approx 1/4)$$
At the same time,
in this case the general unitarity and information are still
retained, i.e. we again have the "unitary" product of two
"nonunitary" arrows:
\begin{eqnarray}
IV.(QMFL,OS,\alpha\approx 1/4)\stackrel{Big\enskip
Bang}{\longrightarrow}(QM,\alpha\approx 0)\stackrel{Big\enskip
Crunch }{\longrightarrow}\nonumber \\(QMFL,FS,\alpha\approx
1/4)\nonumber .
\end{eqnarray}
Finally, arrow III may appear directly, i.e. without the
appearance of arrows I and II, when in the Early Universe mini BH
are arising:
\\
\\
$$IIIb.(QMFL,OS,\alpha\approx 1/4){\longrightarrow}(QMFL,
mini\enskip BH, SBH,\alpha\approx 1/4).$$
\\
\\
Note that here, unlike the previous cases, a unitary transition
occurs immediately, without any additional nonunitary ones, and
with retention of the total information.
\\ Another approach to the information paradox problem associated
with the above-mentioned methods (density matrix deformation) is
the introduction and investigation of a new value, namely entropy
density per minimum unit area. This approach is described in the
following subsection.

\subsection{Entropy density matrix and information loss problem }
In \cite{shalyt1}-\cite{shalyt5} the authors were too careful,
when introducing for density pro-matrix $\rho(\alpha)$ the value
$S_{\alpha}$ generalizing the ordinary statistical entropy:
\\
 $$S_{\alpha}=-Sp[\rho(\alpha)\ln(\rho(\alpha))]=
 -<\ln(\rho(\alpha))>_{\alpha}.$$
\\
In \cite{shalyt4},\cite{shalyt5} it was noted that $S_{\alpha}$
means the entropy density   on a minimum unit area depending on
the scale. In fact a more general concept accepts the form of the
entropy density matrix \cite{shalyt11}:
\begin{equation}\label{U4h}
S^{\alpha_{1}}_{\alpha_{2}}=-Sp[\rho(\alpha_{1})\ln(\rho(\alpha_{2}))]=
-<\ln(\rho(\alpha_{2}))>_{\alpha_{1}},
\end{equation}
where $0< \alpha_{1},\alpha_{2}\leq 1/4.$
\\ $S^{\alpha_{1}}_{\alpha_{2}}$ has a clear physical meaning:
the entropy density is computed  on the scale associated with the
deformation parameter $\alpha_{2}$ by the observer who is at a
scale corresponding to the deformation parameter $\alpha_{1}$.
Note that with this approach the diagonal element
$S_{\alpha}=S_{\alpha}^{\alpha}$,of the described matrix
$S^{\alpha_{1}}_{\alpha_{2}}$ is the density of entropy, measured
by the observer  who is at the same scale  as the measured object
associated with the deformation parameter $\alpha$. In \cite{r15}
and section 4.3 such a construction was used implicitly in
derivation of the semiclassical Bekenstein-Hawking formula for the
Black Hole entropy:

a) for the initial (approximately pure) state
\\
$$S_{in}=S_{0}^{0}=0,$$
\\
b) using the exponential ansatz(\ref{U15}),we obtain:
\\
$$S_{out}=S^{0}_{\frac{1}{4}}=-<ln[exp(-1/4)]\rho_{pure}>=-<\ln(\rho(1/4))>
=\frac{1}{4}.$$
\\
So increase in the entropy density for an external observer at the
large-scale limit is 1/4. Note that increase of the entropy
density(information loss) for the observer crossing the horizon of
the black hole's events and moving with the information flow to
singularity will be smaller:
\begin{eqnarray}
S_{out}=S_{\frac{1}{4}}^{\frac{1}{4}}=-Sp(exp(-1/4)
ln[exp(-1/4)]\rho_{pure}) \nonumber \\
=-<\ln(\rho(1/4))>_{\frac{1}{4}} \approx 0.1947 \nonumber
\end{eqnarray}
It is clear that this fact may be interpreted as follows: for the
observer moving together with information its loss can occur only
at the transition to smaller scales, i.e. to greater deformation
parameter $\alpha$. \\
\\ Now we consider the general Information Problem.
Note that with the well-known Quantum Mechanics (QM) the entropy
density matrix $S^{\alpha_{1}}_{\alpha_{2}}$ (\ref{U4h}) is
reduced only to one element $S_{0}^{0}$ . So we can not test
anything. Moreover, in previous works relating the quantum
mechanics of black holes and information paradox
\cite{r15},\cite{r13},\cite{r16} the initial and final states when
a particle hits the hole are treated proceeding from different
theories (QM and QMFL respectively), as was indicated in diagram
II:
\\
\\
(Large-scale limit, QM,
 density matrix) $\rightarrow$ (Black Hole, singularity, QMFL,
density pro-matrix).
\\
\\
Of course in this case any conservation of information is
impossible as these theories are based on different concepts of
entropy. Simply saying, it is incorrect to compare the entropy
interpretations of two different theories (QM and QMFL)where this
notion is originally differently understood. So the chain above
must be symmetrized by accompaniment of the arrow on the left ,so
in an ordinary situation we have a chain (diagram III):
\\
\\
(Early Universe, origin singularity, QMFL, density pro-matrix)
$\rightarrow$
\\ (Large-scale limit, QM,
 density matrix)$\rightarrow$ (Black Hole, singularity, QMFL,
density pro-matrix).
\\
\\
So it's more correct to compare entropy close to the origin and
final (Black hole) singularities. In other words, it is necessary
to take into account not only the state, where information
disappears, but also that whence it appears. The question arises,
whether in this case the information is lost for every separate
observer. For the event under consideration this question sounds
as follows: are the entropy densities S(in) and S(out) equal for
every separate observer? It will be shown that in all conceivable
cases they are equal.

1) For the observer in the large-scale limit (producing
measurements in the semiclassical approximation) $\alpha_{1}=0$
\\
\\
$S(in)=S^{0}_{\frac{1}{4}}$ (Origin singularity)
\\
\\
$S(out)=S^{0}_{\frac{1}{4}}$ (Singularity in Black Hole)
\\
\\
So $S(in)=S(out)=S^{0}_{\frac{1}{4}}$. Consequently, the initial
and final densities of entropy are equal and there is no
information loss.
\\
2) For the observer moving together with the information flow in
the general situation  we have the chain:
\\
$$S(in)\rightarrow S(large-scale)\rightarrow S(out),$$
\\
where $S(large-scale)=S^{0}_{0}=S$. Here $S$ is an ordinary
entropy of Quantum Mechanics(QM), but
$S(in)=S(out)=S^{\frac{1}{4}}_{\frac{1}{4}}$,- value considered in
QMFL. So in this case the initial and final densities of entropy
are equal without any loss of information.
\\
3) This is a special case of 2), when we do not leave out of the
Early Universe considering the processes with the participation of
black mini-holes only. In this case the originally specified chain
becomes shorter by one section (diagram IIIb):
\\
\\
(Early Universe, origin singularity, QMFL, density
pro-matrix)$\rightarrow$ (Black Mini-Hole, singularity, QMFL,
density pro-matrix),
\\
\\
and member $S(large-scale)=S^{0}_{0}=S$ disappears at the
corresponding chain of the entropy density associated with the
large-scale:
\\
$$S(in)\rightarrow S(out),$$
\\
It is, however, obvious that in case
$S(in)=S(out)=S^{\frac{1}{4}}_{\frac{1}{4}}$ the density of
entropy is preserved. Actually this event was mentioned in
\cite{shalyt5},where from the basic principles it has been found
that black mini-holes do not radiate, just in agreement with the
results of other authors \cite{r17},\cite{r31},\cite{r30}.
\\ As a result, it's possible to write briefly
\\
$$S(in)=S(out)=S^{\alpha}_{\frac{1}{4}},$$
\\
where $\alpha$ - any value in the interval $0<\alpha\leq 1/4.$
\\ Actually our inferences are similar to those of section 4.1
in terms of the Liouville's equation deformation:
\\
$$\frac{d\rho}{dt}=\sum_{i}
\frac{d\omega_{i}[\alpha(t)]}{dt}|i(t)><i(t)|-i[H,\rho(\alpha)]=
\\d[ln\omega(\alpha)]\rho (\alpha)-i[H,\rho(\alpha)].$$
\\
The main result of this section is a necessity to account for the
member $d[ln\omega(\alpha)]\rho (\alpha)$,deforming the right-side
expression of $\alpha\approx 1/4$.
\\It is important to note the following.
\\Different approaches are taken to the information loss problem
and unitarity violation in the black holes. In this section
a problem of R.Penroze is approached first. As indicated
in section 4.2, just R.Penroze in \cite{r19}
has demonstrated that information in the black hole may be lost
and unitarity may be violated because of the singularity. Now in
this section we point \cite{shalyt13s} to a possibility for
solving the problem in principle positively with the use of the
canonical Hawking approach \cite{r15},\cite{r16}, i.e. for {\bf
information losses near the horizon of events} of the black hole.
And independently of the results presented in
\cite{r15},\cite{r16} from the developed formalism it follows that
the state measured near the horizon of events is always mixed. In
\cite{r15},\cite{r16},\cite{r13} this is established in view of
thermal radiation exhibited by the black hole. However, in the
last few years the fact of the existence of such a radiation is
open to dispute (e.g., see \cite{r30}). Therefore, we exclude this
fact from our consideration. Using the developed formalism, one is
enabled to arrive to this result with the above-mentioned entropy
density matrix. Actually, in recent works (e.g.,
\cite{r30},\cite{Hooft}), it has been shown that near the horizon
of events the quantum-gravitational effects are considerable.
Proceeding from the entropy density matrix, this means that for
the matrix element $S^{\alpha_{1}}_{\alpha_{2}}$ we always have
$\alpha_{2}>0$ as the quantum-gravitational effects are affecting
small scales only.
\\Now we analyze a random matrix element $S^{\alpha_{1}}_{\alpha_{2}}$.
Obviously, this element may be zero only for the case when
$\rho(\alpha_{2})$ is a pure state measured in QM, i.e.
$\alpha_{2}\approx 0$. A partial case of this situation has been
considered in \cite{r15}, where $\alpha_{1}=\alpha_{2}\approx 0$
and hence $\rho(\alpha_{1})=\rho(\alpha_{2})=\rho_{in}$ with zero
entropy
\begin{equation}\label{Us5}
S^{0}_{0}=-Sp[\rho_{in}\ln(\rho_{in})]=0.
\end{equation}
Actually, this is the initial state entropy measured in the
original statement of the information paradox problem
\cite{r15},\cite{r16},\cite{r13}
\begin{equation}\label{Us6}
S^{in}=S^{0}_{0}=-Sp[\rho_{in}\ln(\rho_{in})]=
-Sp[\rho_{pure}\ln(\rho_{pure})]=0.
\end{equation}
The question is, what can be measured by an observer at the exit
for $\rho_{out}$, when all measurements are performed in QM, i.e.
when we have at hand only one deformation parameter
$\alpha_{1}\approx 0$? Simply this means that in QMFL due to the
quantum-gravitational effects at a horizon of the above events in
the black hole \cite{r30},\cite{Hooft} leads to:
\begin{equation}\label{Us7}
S^{out}=S^{\alpha_{2}}_{\alpha_{2}}=-Sp[\rho_{out}\ln(\rho_{out})]=
-Sp[\rho_{\alpha_{2}}\ln(\rho_{\alpha_{2}})]\neq 0
\end{equation}
for some $\alpha_{2}>0$,unapproachable in QM, is conforming to a
particular mixed state in QM with the same entropy
\begin{equation}\label{Us8}
S^{out}=-Sp[\rho_{out}\ln(\rho_{out})]=
-Sp[\rho_{mix}\ln(\rho_{mix})]\neq 0.
\end{equation}
It should be emphasized that a mixed state in (\ref{Us8}) will be
not uniquely defined. Of course, in this statement there is
information loss, since
\begin{equation}\label{Us9}
\Delta S=S^{out}-S^{in}>0.
\end{equation}
However, as shown in \cite{shalyt11},\cite{shalyt13}, it will be
more correct to consider the state close to the origin singularity
(or in the early Universe where the quantum-gravitational effects
should be also included) as an initial state for which $S^{in}$ is
calculated, naturally with certain $\alpha>0$ and with entropy
\begin{equation}\label{Us10}
S^{in}=S^{\alpha}_{\alpha}=-Sp[\rho_{in}\ln(\rho_{in})]=
-Sp[\rho_{\alpha}\ln(\rho_{\alpha})]>0.
\end{equation}
Again for the observer making measurements in QM and having no
access to any $\alpha>0$ (i.e. having access to $\alpha\approx 0$
only) this is associated with a certain mixed state
\begin{equation}\label{Us11}
S^{in}=-Sp[\rho_{in}\ln(\rho_{in})]=
-Sp[\rho_{mix}\ln(\rho_{mix})]>0
\end{equation}
that is also ambiguously determined. In this way the
superscattering operator determined in
\cite{r13},\cite{r15},\cite{r16}
\\
$$\$:\rho^{in}\rightarrow\rho^{out}\enskip or
\enskip\$:\rho_{pure}\rightarrow\rho_{mix}$$
\\
in case under consideration will be of the form
\\
$$\$:\rho_{mix}\rightarrow\rho_{mix}.$$
\\
Since in this case $S^{in}>0$ and $S^{out}>0$, may be no
information loss, then
\begin{equation}\label{Us12}
\Delta S=S^{out}-S^{in}=0.
\end{equation}
The following points of particular importance should be taken into
consideration:
\\1) A study of the information paradox problem in the generalized
Quantum Mechanics (QMFL) provides extended possibilities for
interpretation of the notion of entropy. Indeed, in the classical
problem statement \cite{r15},\cite{r13},\cite{r16}
$S^{in}=-Sp[\rho_{in}\ln(\rho_{in})]$ is compared with
$S^{out}=-Sp[\rho_{out}\ln(\rho_{out})]$,i.e. within the scope of
the above-mentioned entropy density $S^{\alpha_{1}}_{\alpha_{2}}$
, introduced in \cite{shalyt11}, \cite{shalyt13}, two different
diagonal elements $S^{0}_{0}$ and $S^{\alpha}_{\alpha}$ are
compared. However, in the paradigm under consideration one is free
to compare $S^{0}_{0}$ to $S^{0}_{\alpha}$ or
$S^{in}_{in}=-Sp[\rho_{in}\ln(\rho_{in})]$ to
$S^{in}_{out}=-Sp[\rho_{in}\ln(\rho_{out})].$
\\
\\2)Proceeding from this observation and from the results
of \cite{shalyt11},\cite{shalyt13}, we come to the conclusion that
the notion of entropy is {\bf relative} within the generalized
Quantum Mechanics (QMFL) in a sense that it is dependent on two
parameters $\alpha_{1}$ and $\alpha_{2}$ characterizing the
positions of observer and observable, respectively. Certainly, as
projected to QM, it becomes absolute since in this case only one
parameter $\alpha\approx 0;$ is measured.
\\
\\3) As indicated above, within the scope of QMFL it is obvious
that the states close to the origin singularity are always mixed,
being associated with the parameter $\alpha>0$ and hence with
nonzero entropy. At the same time, within QM one can have an
understanding (at least heuristically), in what way mixed states
are generated by the origin singularity.  Actually, in the
vicinity of the origin singularity, i.e. at Planck's scale (where
the quantum-gravitational effects are considerable) the space-time
foam is formed \cite{r25},\cite{Foam} that from the
quantum-mechanical viewpoint is capable of generating only a mixed
state, the components of which are associated with metrics from
the space-time foam with certain probabilities arising from the
partition function for quantum gravitation
\cite{r25},\cite{Hawk-f}.
\subsection {Entropy Bounds,Entropy
Density and Holographic Principle}
In the last few years Quantum
Mechanics of black holes has been studied under the assumption
that GUR are valid \cite{r17},\cite{r31}. As a result of this
approach, it is indicated that the evaporation process of a black
hole gives a stable remnant with a mass on the order of the
Planck's $M_{p}$. However, J.Bekenstein in \cite{bek1} has
credited such an approach as problematic, since then the objects
with dimensions on the order of the Planck length $\sim
10^{-33}cm$ should have very great entropy thus making problems in
regard to the entropy bounds of the black hole remnants
\cite{bek2}.
\\ In connection with this remark of J.\,Bekenstein \cite{bek1}
the following points should be emphasized:
\\ I. An approach proposed in \cite{shalyt11},\cite{shalyt13} and in the
present paper gives a deeper insight into the cause of high
entropy for Planck's black hole remnants, namely: high entropy
density that by this approach at Planck scales takes place for
every fixed observer including that on a customary scale, i.e. on
$\alpha\approx 0$. In \cite{shalyt13} using the exponential ansatz
(section 3) it has been demonstrated how this density can increase
in the vicinity of the singularities with
\\
$$S_{in}=S_{0}^{0}\approx 0$$
\\
up to\\ $$S_{out}=S^{0}_{\frac{1}{4}}=-<ln[exp(-1/4)]\rho_{pure}>
=-<\ln(\rho^{*}(1/4))> =\frac{1}{4}.$$
\\
when the initial state measured by the observer is pure.
\\ As demonstrated in \cite{shalyt11},\cite{shalyt13}, increase
in the entropy density  will be realized also for the observer
moving together with the information flow:
$S_{out}=S^{\frac{1}{4}}_{\frac{1}{4}}>S_{0}^{0}$, though to a
lesser extent than in the first case. Obviously, provided the
existing solutions for (\ref{U13}) are different from the
exponential ansatz, the entropy density for them
$S^{0}_{\alpha_{2}}$ will be increasing as compared to $S_{0}^{0}$
with a tendency of $\alpha_{2}$ to 1/4.
\\II. In works of J.Bekenstein, \cite{bek2} in particular,
a "universal entropy bound" has been used \cite{bek3}:
\begin{equation}\label{UBec}
S\leq 2\pi MR/\hbar,
\end{equation}
where $M$ is the total gravitational mass of the matter and $R$ is
the radius of the smallest sphere that barely fits around a
system. This bound is, however, valid for a weakly gravitating
matter system only. In case of black hole remnants under study it
is impossible to assume that on Planck scales we are concerned
with a weakly gravitating matter system, as in this case the
transition to the Planck's energies is realized where
quantum-gravitational effects are appreciable, and within the
proposed paradigm  parameter $\alpha\approx 0$ is changed by the
parameter $\alpha>0$ or equally QM is changed by QMFL.
\\
\\III.This necessitates mentioning of the recent findings of R.Bousso
\cite{bou1},\cite{bou2}, who has derived the Bekenstein's
"universal entropy bound" for a weakly gravitating matter system,
and among other things in flat space, from the covariant entropy
bound \cite{bou3} associated with the holographic principle of
Hooft-Susskind \cite{hol1},\cite{hol2},\cite{hol3}.
\\ Also it should be noted that the approach proposed in
\cite{shalyt13},\cite{shalyt5} and in the present paper is
consistent with the holographic principle \cite{hol1}-\cite{hol3}.
Specifically, with the use of this approach one is enabled to
obtain the entropy bounds for nonblack hole objects of L.Susskind
\cite{hol2}. Of course, in (\cite{shalyt5}, section 6) and
(\cite{shalyt13}, section 4) it has been demonstrated, how a
well-known semiclassical Bekenstein-Hawking formula for black hole
entropy may be obtained using the proposed paradigm. Then we can
resort to reasoning from \cite{hol2}: "using gedanken experiment,
take a neutral non-rotating spherical object containing entropy
$S$ which fits entirely inside a spherical surface of the area
$A$, and it to collapse to black hole". Whence
\begin{equation}\label{USuss}
S\leq \frac{A}{4l^{2}_{p}}.
\end{equation}
Note also that the entropy density matrix
$S^{\alpha_{1}}_{\alpha_{2}}$ by its definition
\cite{shalyt11},\cite{shalyt13} falls into 2D objects, being
associated with $l^{2}_{min}\sim l^{2}_{p}$ \cite{shalyt5} and
hence implicitly pointing to the holographic principle.
\\Qualitative analysis
performed in this work reveals that the Information Loss Problem
in black holes with the canonical problem statement
 \cite{r15},\cite{r13},\cite{r16} suggests in principle positive solution
within the scope of the proposed method - high-energy density
matrix deformation. Actually, this problem necessitates further
(now quantitative) analysis. Besides, it is interesting to find
direct relations between the described methods and the holographic
principle. Of particular importance seems a conjecture following
from \cite{bou2}:
\\is it possible to derive GUR for high energies
(of strong gravitational field) with the use of the covariant
entropy bound \cite{bou3} in much the same manner as R.Bousso
\cite{bou2} has developed the Heisenberg uncertainty principle for
the flat space?

\subsection {Unitarity, non-unitarity and Heisenberg's algebra
deformation}
The above-mentioned unitary and nonunitary
transitions may be described in terms of Heisenberg's algebra
deformation (deformation of commutators) as well. We use the
principal results and designations from \cite{Magg}.In the process
the following assumptions are resultant: 1)The three-dimensional
rotation group is not deformed; angular momentum ${\bf J}$
satisfies the undeformed $SU(2)$ commutation relations, whereas
the coordinate and momenta satisfy the undeformed commutation
relations $\left[ J_i,x_j\right] =i\epsilon_{ijk}x_k, \left[
J_i,p_j\right] =i\epsilon_{ijk}p_k$. 2) The momenta commute
between themselves: $\left[ p_i,p_j\right] =0$, so the translation
group is also not deformed. 3) Commutators $\left[ x,x\right]$ and
$\left[ x,p\right]$ depend on the deformation parameter $\kappa$
with the dimension of mass. In the limit $\kappa\rightarrow
\infty$ with $\kappa$ much larger than any energy the canonical
commutation relations are recovered.
\\
For a specific realization of points 1) to 3) the generating GUR
are of the form \cite{Magg}: ($\kappa$-deformed Heisenberg
algebra)
\begin{eqnarray}
\left[ x_i ,x_j \right] &= & -\frac{\hbar^2}{\kappa^2}\,
i\epsilon_{ijk}J_k\label{xx}\\ \left[ x_i , p_j \right]   &= &
i\hbar\delta_{ij} (1+\frac{E^2}{\kappa^2})^{1/2}\, .\label{xp}
\end{eqnarray}
Here $E^2=p^2+m^2$. Note that in this formalism the transition
from GUR to UR, or equally from QMFL to QM with $\kappa\rightarrow
\infty$ or from Planck scale to the conventional one, is
nonunitary exactly following the transition from density
pro-matrix to the density matrix in previous sections:
\\
$$\rho(\alpha\neq 0)\stackrel{\alpha\rightarrow
0}{\longrightarrow}\rho.$$
\\
Then the first arrow I in the formalism of this section may be as
follows:
\\
$$I^{\prime}.(GUR,OS,\kappa\sim M_{p})\stackrel{Big\enskip
Bang}{\longrightarrow}(UR,\kappa=\infty)$$ or what is the same
$$I^{\prime\prime}.(QMFL,OS,\kappa\sim M_{p})\stackrel{Big\enskip
Bang}{\longrightarrow}(QM,\kappa=\infty),$$
\\
where $M_{p}$ is the Planck mass.
\\
\\In some works of the last two
years Quantum Mechanics for a Black Hole has been already
considered as a Quantum Mechanics with GUR \cite{r17},\cite{r31}.
As a consequence, by this approach the Black Hole is not
completely evaporated but rather some stable remnants always
remain in the process of its evaporation with a mass $\sim M_{p}$.
In terms of \cite{Magg} this means nothing else but a reverse
transition: $(\kappa=\infty)\rightarrow(\kappa\sim M_{p})$. And
for an outside observer this transition is of the form:
\\
$$II^{\prime}.(UR,\kappa=\infty)\stackrel{absorbing\enskip
BH}{\longrightarrow}(GUR,SBH,\kappa\sim M_{p}),$$ that is
$$II^{\prime\prime}.(QM,\kappa=\infty)\stackrel{absorbing\enskip
BH}{\longrightarrow}(QMFL,SBH,\kappa\sim M_{p}).$$
\\
\\
So similar to the previous section, two nonunitary inverse
transitions a)$I^{\prime},(I^{\prime\prime})$ and
b)$II^{\prime},(II^{\prime\prime})$ are liable to generate a
unitary transition:
\begin{eqnarray}
III^{\prime}.(GUR,OS,\kappa\sim M_{p})\stackrel{Big\enskip
Bang}{\longrightarrow}\nonumber \\
(UR,\kappa=\infty)\stackrel{absorbing\enskip
BH}{\longrightarrow}(GUR,SBH,\kappa\sim M_{p}) \nonumber
\end{eqnarray} or
to summerize
\\
$$III^{\prime\prime}.(GUR,OS,\kappa\sim
M_{p})\rightarrow(GUR,SBH,\kappa\sim M_{p})$$
\\
In conclusion of this section it should be noted that an approach
to the Quantum Mechanics at Planck Scale using the Heisenberg
algebra deformation (similar to the approach based on the density
matrix deformation from the  section3) gives a deeper insight into
the possibility of retaining the unitarity and the total quantity
of information in the Universe, making possible the solution of
Hawking's Information Paradox Problem
\cite{r15},\cite{r13},\cite{r16}.

\subsection {Statistical mechanics deformation and transitions}
Naturally, deformation of Quantum Mechanics in the Early Universe
is associated with the Statistical Mechanics deformation as
indicated in \cite{shalyt6,shalyt7}. In case under consideration
this simply implies a transition from the Generalized Uncertainty
Relations (GUR) of Quantum Mechanics to GUR in Thermodynamics
\cite{shalyt7},\cite{shalyt9,shalyt10}. The latter are
distinguished from the normal uncertainty relations by:
\begin{equation}\label{U1T}
\Delta \frac{1}{T}\geq\frac{k}{\Delta U},
\end{equation}
i.e. by inclusion of the high-temperature term into the right-hand
side (section 6 of this chapter)
\begin{equation}\label{U2T}
\Delta \frac{1}{T}\geq
  \frac{k}{\Delta U}+\alpha^{\prime}
  \frac{1}{T_{p}^2}\frac{\Delta U}{k}+...     .
\end{equation}
Thus, denoting the Generalized Uncertainty Relations in
Thermodynamics as GURT and using abbreviation URT for the
conventional ones, we obtain a new form of diagram I from section
III ($I^{\prime}$ of section IV respectively):
\\
 $$I^{T}.(GURT,OS)\stackrel{Big\enskip
Bang}{\longrightarrow}(URT).$$
\\
In \cite{shalyt6,shalyt7} and section 5 of this chapter the
Statistical Mechanics deformation associated with GURT is
implicitly assumed by the introduction of the respective
deformation for the statistical density matrix $\rho_{stat}(\tau)$
where $0<\tau \leq 1/4$. Obviously, close to the Origin
Singularity $\tau\approx 1/4$. Because of this, arrow $I^{T}$ may
be represented in a more general form as
\\
 $$I^{Stat}.(GURT,OS,\rho_{stat}(\tau),\tau\approx 1/4)
 \stackrel{Big\enskip
Bang}{\longrightarrow}(URT,\rho_{stat},\tau\approx 0).$$
\\
The reverse transition is also possible. In \cite{r17},\cite{r31}
it has bee shown that Statistical Mechanics of Black Hole should
be consistent with the deformation of a well-known Statistical
Mechanics. The demonstration of an *upper* bound for temperature
in Nature, given by Planck temperature and related to Black Hole
evaporation, was provided in \cite{Castro3}. It is clear that
emergence of such a high temperatures is due to GURT. And we have
the following diagram that is an analog of diagrams II and
$II^{\prime}$ for Statistical Mechanics:
\\
$$II^{Stat}.(URT,\rho_{stat},\tau\approx 0)
\stackrel{absorbing\enskip
BH}{\longrightarrow}(GURT,SBH,\rho_{stat}(\tau),\tau\approx 1/4).
$$
\\
By this means, combining $I^{Stat}$ and $II^{Stat}$, we obtain
$III^{Stat}$ representing a complete statistical-mechanics analog
for quantum-mechanics diagrams $III$ and $III^{\prime}$:
\begin{eqnarray}
III^{Stat}.(GURT,OS,\tau\approx 1/4)
 \stackrel{Big\enskip
Bang,\enskip absorbing\enskip BH}{\longrightarrow} \nonumber \\
(GURT, SBH,\tau\approx 1/4).\nonumber
\end{eqnarray}
And in this case two nonunitary transitions $I^{Stat}$ and
$II^{Stat}$ in the end lead to a unitary transition $III^{Stat}$.

\section{The Universe as a Nonuniform Lattice in Finite-Volume
Hypercube}
 In this section a new small parameter associated with
the density matrix deformation (density pro-matrix)studied in
previous sections  is introduced into the Generalized Quantum
Mechanics (GQM), i.e. quantum mechanics involving description of
the Early Universe. It is noted that this parameter has its
counterpart in the Generalized Statistical Mechanics. Both
parameters offer a number of merits: they are dimensionless,
varying over the interval from 0 to 1/4 and assuming in this
interval a discrete series of values. Besides, their definitions
contain all the fundamental constants. These parameters are very
small for the conventional scales and temperatures, e.g. the value
of the first parameter is on the order of $\approx 10^{-66+2n}$,
where $10^{-n}$ is the measuring scale and the Planck scale $\sim
10^{-33}cm$ is assumed. The second one is also too small for the
conventional temperatures, that is those much below the Planck's.
It is demonstrated that relative to the first of these parameters
the Universe may be considered as a nonuniform lattice in the
four-dimensional hypercube with dimensionless finite-length (1/4)
edges. And the time variable is also described by one of the
above-mentioned dimensions due to the second parameter and
Generalized Uncertainty Relation in thermodynamics. In this
context the lattice is understood as a deformation rather than
approximation \cite{shalyt14}.

\subsection{Definition of lattice}
It should be noted that according to subsection 3.2 a minimum
measurable length is equal
 to $l^{*}_{min}=2l_{min}$ being a nonreal
number at point $l_{min}$,$Sp[\rho(\alpha)]$. Because of this, a
space part of the Universe is a lattice with a spacing of
$a_{min}=2l_{min}\sim 2l_{p}$. In consequence the first issue
concerns the lattice spacing of any lattice-type model(for example
\cite{rLat1,rLat2}): a selected lattice spacing $a_{lat}$ should
not be less than $a_{min}$,i.e. always $a_{lat}\geq a_{min}>0$.
Besides, a continuum limit in any lattice-type model is meaning
$a_{lat}\rightarrow a_{min}>0$ rather than $a_{lat}\rightarrow 0$.
\\ Proceeding from $\alpha$, for each space dimension we have a
discrete series of rational values for the inverse squares of even
numbers nonuniformly distributed along the real number line
$\alpha = 1/4, 1/16,1/36,1/64,...$. A question arises,is this
series somewhere terminated or, on the contrary, is it infinite?
The answer depends on the answers to two other questions:
\\(1) Is there theoretically a maximum measurability limit for the scales
$l_{max}$?  and
\\(2) Is our Universe closed in the sense that its
extension may be sometime replaced by compression, when a maximum
extension precisely gives a maximum scale $l_{max}$?
\\Should an answer to one of these questions be positive, we should have
$0<l^{2}_{min}/l^{2}_{max}\leq\alpha\leq1/4$ rather than condition
1 of {\bf Definition 1, subsection 3.1 of this chapter}
\\ Note that in the majority of cases all three space dimensions are
equal, at least at large scales, and hence their associated values
of $\alpha$ parameter should be identical. This means that for
most cases, at any rate in the large-scale (low-energy) limit, a
single deformation parameter $\alpha$ is sufficient to accept one
and the same value for all three dimensions to a high degree of
accuracy. In the general case, however, this is not true, at least
for very high energies (on the order of the Planck's), i.e. at
Planck scales, due to noncommutativity of the spatial coordinates
\cite{r4},\cite{Magg}:
\\
$$\left[ x_i ,x_j \right]\neq 0.$$
\\
In consequence in the general case we have a point with
coordinates ${\bf
\widetilde{\alpha}}=(\alpha_{1},\alpha_{2},\alpha_{3})$ in the
normal(three-dimensional) cube $I_{1/4}^{3}$ of side
$I_{1/4}=(0;1/4]$.
\\ It should be noted that this universal cube may be extended to
the four-dimensional hypercube by inclusion of the additional
parameter $\tau,\tau\in I_{1/4}$ that is generated by internal
energy of the statistical ensemble and its temperature for the
events when this notion is the case. It will be recalled that
$\tau$ parameter occurs from a maximum temperature that is  in its
turn generated by the Generalized Uncertainty Relations of "energy
- time" pair in GUR (see {\bf Definition 2} in subsection 5.1 and
\cite{shalyt6,shalyt7}).
\\So $\tau$ is a counterpart (twin) of $\alpha$, yet for the Statistical
Mechanics. At the same time, originally for $\tau$ nothing implies
the discrete properties of parameter $\alpha$ indicated above:
\\ for $\tau$ there is a discrete series (lattice) of the rational
values of inverse squares for even numbers not uniformly
distributed along the real number line: $\tau = 1/4, 1/16,
1/36,1/64,...$.
 \\ Provided such a series exists actually,
\\The finitness and infinity question for this series amounts to
two other questions:
\\(1) Is there theoretically any minimum measurability limit for
the average temperature of the Universe $T_{min}\neq 0$ and
\\(2) Is our Universe closed in a sense that its extension may be sometime
replaced by compression? Then maximum extension just gives a
minimum temperature $T_{min}\neq 0$.
\\ The question concerning the discretization of parameter $\tau$
is far from being idle. The point is that originally by its nature
this parameter seems to be continuous as it is associated with
temperature. Nevertheless, in the following section we show that
actually $\tau$ is dual in nature: it is directly related to time
that is in turn quantized,in the end giving a series $\tau = 1/4,
1/16, 1/36,1/64,...$.

\subsection{Dual nature of parameter $\tau$ and its temporal
aspect}

In this way when at point ${\bf\widetilde{\alpha}}$ of the normal
(three-dimensional) cube $I_{1/4}^{3}$ of side $I_{1/4}=(0;1/4]$
an additional "temperature" variable $\tau$ is added, a nonuniform
lattice of the point results, where we denote
$\widetilde{\alpha}_{\tau}=(\widetilde{\alpha},\tau)=
(\alpha_{1},\alpha_{2},\alpha_{3},\tau)$ at the four-dimensional
hypercube $I_{1/4}^{4}$, every coordinate of which assumes one and
the same discrete series of values: 1/4, 1/16, 1/36,1/64,...,
$1/4n^{2}$,... .(Further it is demonstrated that $\tau$ is also
taking on a discrete series of values.) The question arises,
whether time "falls" within this picture. The answer is positive.
Indeed, parameter $\tau$ is dual (thermal and temporal) in nature
owing to introduction of the Generalized Uncertainty Relations in
Thermodynamics (GURT)
(\cite{shalyt9},\cite{shalyt10},\cite{shalyt7} and section 6):
\\
$$\Delta \frac{1}{T} \geq
  \frac{k}{\Delta U}+\alpha^{\prime}
  \frac{1}{T_{p}^2}\,
  \frac{\Delta U}{k}+...,$$
\\
where $k$ - Boltzmann constant, $T$ - temperature of the ensemble,
$U$ - its internal energy. A direct implication of the latter
inequality is occurrence of a "maximum" temperature $T_{max}$ that
is inversely proportional to "minimal" time $t_{min}\sim t_{p}$:
\\
$$T_{max}=\frac{\hbar}{2\surd \alpha^{\prime}t_{p}
k}=\frac{\hbar}{\Delta t_{min} k}$$
\\
However, $t_{min}$ follows from the Generalized Uncertainty
Relations in Quantum Mechanics for "energy-time" pair
(\cite{shalyt6},\cite{shalyt7} and section 5):
\\
$$\Delta t\geq\frac{\hbar}{\Delta
E}+\alpha^{\prime}t_{p}^2\,\frac{\Delta E}{ \hbar}.$$
\\
Thus, $T_{max}$ is the value relating GUR and GURT together (see
sections 5,6 and \cite{shalyt9},\cite{shalyt10},\cite{shalyt7})
\begin{equation}\label{U18L}
\left\{
\begin{array}{lll}
\Delta x & \geq & \frac{\displaystyle\hbar}{\displaystyle\Delta
p}+\alpha^{\prime} L_{p}^2\,\frac{\displaystyle\Delta
p}{\displaystyle\hbar}+... \\
  &  &  \\
  \Delta t & \geq &  \frac{\displaystyle\hbar}{\displaystyle\Delta E}+\alpha^{\prime}
  t_{p}^2\,\frac{\displaystyle\Delta E}{\displaystyle\hbar}+... \\
  &  &  \\

  \Delta \frac{\displaystyle 1}{\displaystyle T} & \geq &
  \frac{\displaystyle k}{\displaystyle\Delta U}+\alpha^{\prime}
  \frac{\displaystyle 1}{\displaystyle T_{p}^2}\,
  \frac{\displaystyle\Delta U}{\displaystyle k}+...,
\end{array} \right.
\end{equation},
since the thermodynamic value $T_{max}$ (GURT) is associated with
the quantum-mechanical one $E_{max}$ (GUR) by the formula from
section 5:
\\
$$T_{max}=\frac{E_{max}}{k}$$
\\
The notion of value $t_{min}\sim 1/T_{max}$ is physically crystal
clear, it means a minimum time for which any variations in the
energy spectrum of every physical system may be recorded.
Actually, this value is equal to $t^{*}_{min}=2t_{min}\sim t_{p}$
as at the initial points $l_{min}$ and $T_{max}$ the spurs of the
quantum-mechanical and statistical density pro-matrices
${\bf\rho_(\alpha)}$ and ${\bf\rho_{stat}(\tau)}$ are complex,
determined only beginning from $2l_{min}$ and
$T^{*}_{max}=\frac{1}{2}T_{max}$ \cite{shalyt5},\cite{shalyt7}
that is associated with the same time point
$t^{*}_{min}=2t_{min}$. For QMFL this has been noted in the
previous section.
\\  In such a manner a discrete series $l^{*}_{min},2l^{*}_{min}$,...
generates in QMFL the discrete time series
$t^{*}_{min},2t^{*}_{min},...$, that is in turn associated (due to
GURT)with a discrete temperature series
$T^{*}_{max}$,$\frac{1}{2}T^{*}_{max}$, ... . From this it is
inferred that a "temperature" scale $\tau$ may be interpreted as a
"temporal" one $\tau=t_{min}^{2}/t^{2}$. In both cases the
generated series has one and the same discrete set of values for
parameter $\tau$ :$\tau = 1/4, 1/16, 1/36,1/64,..., 1/4n^{2}$,...
.. Thus, owing to time quantization in QMFL, one is enabled to
realize quantization of temperature in the generalized Statistical
Mechanics with the use of GURT.
\\Using $Lat_{\widetilde{\alpha}}$, we denote the lattice in cube
$I_{1/4}^{3}$ formed by points $\widetilde{\alpha}$, and through
$Lat^{\tau}_{\widetilde{\alpha}}$ we denote the lattice in
hypercube $I_{1/4}^{4}$ that is formed by points
$\widetilde{\alpha}_{\tau}=(\widetilde{\alpha},\tau)$.

\subsection{Quantum theory \\ for the lattice in hypercube}
Any quantum theory may be defined for the indicated lattice in
hypercube.
 To this end, we recall the principal result of subsection 4.4 as
{\bf Definition $1^{\prime}$} in this section with $\alpha$
changed by $\widetilde{\alpha}$:
\\ \noindent {\bf Definition
$1^{\prime}$} {\bf Quantum Mechanics with Fundamental Length}
\\ {\bf (Shr{\"o}dinger's picture)}
\\
Here, the prototype of Quantum Mechanical normed wave function (or
the pure state prototype) $\psi(q)$ with $\int|\psi(q)|^{2}dq=1$
in QMFL is
$\psi(\widetilde{\alpha},q)=\theta(\widetilde{\alpha})\psi(q)$.
The parameter of deformation $\widetilde{\alpha}\in I_{1/4}^{3}$.
Its properties are
$|\theta(\widetilde{\alpha})|^{2}<1$,$\lim\limits_{|\widetilde{\alpha}|\rightarrow
0}|\theta(\widetilde{\alpha})|^{2}=1$ and the relation
$|\theta(\alpha_{i})|^{2}-|\theta(\alpha_{i})|^{4}\approx
\alpha_{i}$ takes place. In such a way the total probability
always is less than 1:
$p(\widetilde{\alpha})=|\theta(\widetilde{\alpha})|^{2}
=\int|\theta(\widetilde{\alpha})|^{2}|\psi(q)|^{2}dq<1$ tending to
1, when  $\|\widetilde{\alpha}\|\rightarrow 0$. In the most
general case of the arbitrarily normed state in QMFL(mixed state
prototype)
$\psi=\psi(\widetilde{\alpha},q)=\sum_{n}a_{n}\theta_{n}(\widetilde{\alpha})\psi_{n}(q)$
with $\sum_{n}|a_{n}|^{2}=1$ the total probability is
$p(\widetilde{\alpha})=\sum_{n}|a_{n}|^{2}|\theta_{n}(\widetilde{\alpha})|^{2}<1$
and
 $\lim\limits_{\|\widetilde{\alpha}\|\rightarrow 0}p(\widetilde{\alpha})=1$.

It is natural that Shr{\"o}dinger equation is also deformed in
QMFL. It is replaced by the equation

\begin{equation}\label{U24L}
\frac{\partial\psi(\widetilde{\alpha},q)}{\partial t}
=\frac{\partial[\theta(\widetilde{\alpha})\psi(q)]}{\partial
t}=\frac{\partial\theta(\widetilde{\alpha})}{\partial
t}\psi(q)+\theta(\widetilde{\alpha})\frac{\partial\psi(q)}{\partial
t},
\end{equation}
where the second term in the right-hand side generates the
Shr{\"o}dinger equation as
\begin{equation}\label{U25L}
\theta(\widetilde{\alpha})\frac{\partial\psi(q)}{\partial
t}=\frac{-i\theta(\widetilde{\alpha})}{\hbar}H\psi(q).
\end{equation}
Here $H$ is the Hamiltonian and the first member is added
similarly to the member that appears in the deformed Liouville
equation, vanishing when $\theta[\widetilde{\alpha}(t)]\approx
const$. In particular, this takes place in the low energy limit in
QM, when $\|\widetilde{\alpha}\|\rightarrow 0$. It should be noted
that the above theory is not a time reversal of QM because the
combination $\theta(\widetilde{\alpha})\psi(q)$ breaks down this
property in the deformed Shr{\"o}dinger equation. Time-reversal is
conserved only in the low energy limit, when a quantum mechanical
Shr{\"o}dinger equation is valid.
\\ According to {\bf Definition $1^{\prime}$}everywhere $q$ is
the coordinate of a point at the three-dimensional space. As
indicated in \cite{shalyt1}--\cite{shalyt5} and section 3.2, for a
density pro-matrix there exists an exponential ansatz satisfying
the formula (\ref{U13}) of {\bf Definition 1}, section 3.1:
\begin{equation}\label{U26L}
\rho^{*}(\alpha)=\sum_{i}\omega_{i} exp(-\alpha)|i><i|,
\end{equation}
where all $\omega_{i}>0$ are independent of $\alpha$  and their
sum is equal to 1. In this way
$Sp[\rho^{*}(\alpha)]=exp(-\alpha)$. Then in the momentum
representation $\alpha=p^{2}/p^{2}_{max}$, $p_{max}\sim
p_{pl}$,where $p_{pl}$ is the Planck momentum. When present in
matrix elements, $exp(-\alpha)$  damps the contribution of great
momenta in a perturbation theory.
\\ It is clear that for each of the coordinates $q_{i}$ the
exponential ansatz makes a contribution to the deformed wave
function $\psi(\widetilde{\alpha},q)$ the modulus of which equals
$exp(-\alpha_{i}/2)$  and, obviously, the same contribution to the
conjugate function $\psi^{*}(\widetilde{\alpha},q)$. Because of
this, for exponential ansatz one may write
\begin{equation}\label{U27L}
\psi(\widetilde{\alpha},q)=\theta(\widetilde{\alpha})\psi(q),
\end{equation}
where $|\theta(\widetilde{\alpha})|=exp(-\sum_{i}\alpha_{i}/2)$.
As noted above, the last exponent of the momentum representation
reads $exp(-\sum_{i}p_{i}^{2}/2p_{max}^{2})$ and in this way it
removes UV (ultra-violet) divergences in the theory. It follows
that $\widetilde{\alpha}$ is a new small parameter. Among its
obvious advantages one could name:
\\1)  its dimensionless nature,
\\2)  its variability over the finite interval $0<\alpha_{i}\leq 1/4$.
Besides, for the well-known physics it is actually very small:
$\alpha\sim 10^{-66+2n}$, where $10^{-n}$ is the measuring scale.
Here the Planck scale $\sim 10^{-33}cm$ is assumed;
\\3)and finally the calculation of this parameter involves all
three fundamental constants, since by Definition 1 of subsection
3.1 $\alpha_{i}= l_{min}^{2}/x_{i}^{2 }$, where $x_{i}$ is the
measuring scale on i-coordinate and $l_{min}^{2}\sim
l_{pl}^{2}=G\hbar/c^{3}$.
\\ Therefore, series expansion in $\alpha_{i}$ may be of great importance.
Since all the field components and hence the Lagrangian will be
dependent on $\widetilde{\alpha}$, i.e.
$\psi=\psi(\widetilde{\alpha}),L=L(\widetilde{\alpha})$, quantum
theory may be considered as a theory of lattice
$Lat_{\widetilde{\alpha}}$ and hence of lattice
$Lat^{\tau}_{\widetilde{\alpha}}$.
\subsection{Introduction of
quantum field theory and initial analysis} With the use of this
approach for the customary energies a Quantum Field Theory (QFT)
is introduced with a high degree of accuracy. In our context
"customary" means the energies much lower than the Planck ones.
\\ It is important that as the spacing of lattice
 $Lat^{\tau}_{\widetilde{\alpha}}$
is decreasing in inverse proportion to the square of the
respective node, for a fairly large node number $N>N_{0}$  the
lattice edge beginning at this node $\ell_{N,N+1}$
\cite{shalyt1}--\cite{shalyt5} will be of length $\ell_{N,N+1}\sim
1/4N^{3}$, and by this means edge lengths of the lattice are
rapidly decreasing with the spacing number. Note that in the
large-scale limit this (within any preset accuracy)leads to
parameter $\alpha=0$, pure states and in the end to QFT. In this
way a theory for the above-described lattice presents a
deformation of the originally continuous variant of this theory as
within the developed approach continuity is accurate to $\approx
10^{-66+2n}$, where $10^{-n}$ is the measuring scale and the
Planck scale $\sim 10^{-33}cm$ is assumed. Whereas the lattice per
se $Lat^{\tau}_{\widetilde{\alpha}}$ may be interpreted as a
deformation of the space continuum with the deformation parameter
equal to the varying edge length
$\ell_{\alpha^{1}_{\tau_{1}},\alpha^{2}_{\tau_{2}}}$, where
$\alpha^{1}_{\tau_{1}}$ and $\alpha^{2}_{\tau_{2}}$ are two
adjacent points of the lattice $Lat^{\tau}_{\widetilde{\alpha}}$.
Proceeding from this, all well-known theories including
$\varphi^{4}$, QED, QCD and so on may be studied based on the
above-described lattice.
\\ Here it is expedient to make the following remarks:
\\{\bf (1) going on from the well-known energies of these theories
to higher energies (UV behavior) means a change from description
of the theory's behavior for the lattice portion with high edge
numbers to the portion with low numbers of the edges;
\\ (2) finding of quantum correction factors for the primary deformation
parameter $\widetilde{\alpha}$ is a power series expansion in each
$\alpha_{i}$. In particular, in the simplest case (Definition
$1^{\prime}$ of subsection 8.3 )means expansion of the left side
in relation
$|\theta(\alpha_{i})|^{2}-|\theta(\alpha_{i})|^{4}\approx
\alpha_{i}$:
\\
$$|\theta(\alpha_{i})|^{2}-|\theta(\alpha_{i})|^{4}=
\alpha_{i}+a_{0}\alpha^{2}_{i}+a_{1}\alpha^{3}_{i}+...$$
\\
and calculation of the associated coefficients $a_{0},a_{1},...$.}
This approach to calculation of the quantum correction factors may
be used in the formalism for density pro-matrix (Definition 1 of
subsection 3.1). In this case, the primary relation (\ref{U13}) of
{\bf Definition 1}, section 3.1 may be written in the form of a
series
\begin{equation}\label{U28L}
Sp[\rho(\alpha)]-Sp^{2}[\rho(\alpha)]=\alpha+a_{0}\alpha^{2}
+a_{1}\alpha^{3}+...     .
\end{equation}
As a result, a measurement procedure using the exponential ansatz
 may be understood as the calculation of factors
$a_{0}$,$a_{1}$,... or the definition of additional members in the
exponent "destroying" $a_{0}$,$a_{1}$,... \cite{shalyt13}. It is
easy to check that the exponential ansatz gives $a_{0}=-3/2$,
being coincident with the logarithmic correction factor for the
Black Hole entropy \cite{r22}.
\\ Most often a quantum theory is considered at zero temperature
$T=0$, in this context amounting to nesting of the
three-dimensional lattice $Lat_{\widetilde{\alpha}}$ into the
four-dimensional one:
$Lat^{\tau}_{\widetilde{\alpha}}$:$Lat_{\widetilde{\alpha}}\subset
Lat^{\tau}_{\widetilde{\alpha}}$ and nesting of the cube
$I_{1/4}^{3}$ into the hypercube $I_{1/4}^{4}$ as a bound given by
equation $\tau=0$. However, in the most general case the points
with nonzero values of $\tau$ may be important as there is a
possibility for nonzero temperature $T\neq0$ (quantum field theory
at finite temperature) that is related to the value of $\tau$
parameter, though very small but still nonzero: $\tau\neq0$. To
illustrate: in QCD for the normal lattice \cite{Di} a critical
temperature $T_{c}$ exists so that the following is fulfilled:
\\
at $$T<T_{c}$$ the confinement phase occurs,
\\
and for $$T>T_{c}$$ the deconfinement is the case.
\\ A critical temperature $T_{c}$ is associated with the "critical"
parameter $\tau_{c}=T^{2}_{c}/T^{2}_{max}$  and the selected bound
of hypercube $I_{1/4}^{4}$ set by equation $\tau=\tau_{c}>0$.
\section{Spontaneous Symmetry Breakdown and Restoration
in a Model with Scalar Fields} This section presents applications
of the previous results and
\cite{shalyt4},\cite{shalyt5},\cite{shalyt14} to the symmetry
breaking and restoration problem, using so far simple models with
scalar fields common in cosmology \cite{cosm1},\cite{cosm3}. In
other words, the author's interest is focused at spontaneous
breaking and restoration of the symmetry in $\alpha$-deformations
of the associated theories. And $\alpha$-deformation of wave
functions, i.e. fields, in QFT is of the following form (section
4):
\begin{equation}\label{U2.6I}
\psi(x) \mapsto \psi(\alpha,x)=\mid \theta (\alpha)\mid \psi(x).
\end{equation}
Here
\begin{equation}\label{U2.7I}
\mid \theta (\alpha)\mid =
exp(-\alpha/2)
\end{equation}
or
\begin{equation}\label{U2.8I}
\theta (\alpha)=\pm exp(-\alpha/2)(cos\gamma \pm isin\gamma).
\end{equation}
\\Further in the text (section 9) it is assumed that
\\
\\1) $\alpha$ of the exponential factor in formula (\ref{U2.7I})
is the same for all space coordinates
$\alpha_{1},\alpha_{2},\alpha_{3}$ of a point of lattice
$Lat^{\tau}_{\widetilde{\alpha}}$. This means that as yet the
noncommutativity effect is of no special importance, and parameter
$\alpha$ is determined by the corresponding energy scale;
\\
\\2) parameter $\alpha$ is dependent on time only
$\alpha=\alpha(t)$. This condition is quite natural since $\alpha$
plays a part of the scale factor and is most often dependent
solely on time(especially in cosmology \cite{cosm1},\cite{cosm2});
\\
\\3) as all physical results should be independent
of a selected normalization $\theta (\alpha)$, subsequent choice
of the normalization will be
\begin{equation}\label{U2.9I}
\theta (\alpha)=exp(-\alpha/2),
\end{equation}
that is $\gamma=0$.
\\First, take a well-known Lagrangian for a
scalar field \cite{cosm3}:
\begin{equation}\label{U3.1I}
L=\frac{1}{2}(\partial_{\mu}\phi)^{2}+\frac{1}{2}\mu^{2}\phi^{2}
-\frac{1}{4}\lambda\phi^{4},
\end{equation}
where $\lambda>0$.
\\ Because of the transformation in the above deformation
\begin{equation}\label{U3.2I}
\phi \Rightarrow \phi(\alpha)=exp(-\frac{\alpha}{2})\phi,
\end{equation}
where $\alpha=\alpha(t)$, $\alpha$ is a deformed Lagrangian of the
form
\begin{equation}\label{U3.3I}
L(\alpha)=\frac{1}{2}(\partial_{\mu}\phi(\alpha))^{2}
+\frac{1}{2}\mu^{2}\phi(\alpha)^{2}
-\frac{1}{4}\lambda\phi(\alpha)^{4}.
\end{equation}
It should be noted that a change from Lagrangian to the
Hamiltonian formalism in this case is completely standard
\cite{Div1} with a natural changing of $\phi$ by $\phi(\alpha)$.
Consequently, with the use of a well-known formula
\begin{equation}\label{U3.4I}
H(\alpha)=p\dot{q}-L(\alpha)
\end{equation}
we obtain
\begin{equation}\label{U3.5I}
H(\alpha)=\frac{1}{2}(\partial_{0}\phi(\alpha))^{2}
+\frac{1}{2}\sum_{i=1}^3 (\partial_{i}\phi(\alpha))^{2}
-\frac{1}{2}\mu^{2}\phi(\alpha)^{2}
+\frac{1}{4}\lambda\phi(\alpha)^{4}.
\end{equation}
We write the right-hand side of (\ref{U3.5I}) so as to solve it
for $\phi$ and $\alpha$. As $\alpha$ is depending solely on time,
the only nontrivial term (to within the factor of 1/2) in the
right-hand side of (\ref{U3.5I}) is the following:
\\
$$(\partial_{0}\phi(\alpha))^{2}=
exp(-\alpha)(\frac{1}{4}\dot{\alpha}^{2}\phi^{2}
-\dot{\alpha}\phi\partial_{0}\phi+(\partial_{0}\phi)^{2})$$.
\\
Thus, $\alpha$ - deformed Hamiltonian $H(\alpha)$ is rewritten as
\begin{equation} \label{U3.6I}
H(\alpha)=exp(-\alpha)[-\frac{1}{2}\dot{\alpha}\phi\partial_{0}\phi
+\frac{1}{2}\sum_{j=0}^3 (\partial_{j}\phi)^{2} \nonumber
\\ +(\frac{1}{8}\dot{\alpha}^{2}-\frac{1}{2}\mu^{2})\phi^{2}
+\frac{1}{4}exp(-\alpha)\lambda\phi^{4}].
\end{equation}
To find a minimum of $H(\alpha)$, we make its partial derivatives
with respect $\phi$ and $\alpha$ equal to zero
\begin{equation}\label{U3.7I}
\left\{
\begin{array}{ll}
\frac{\displaystyle\partial H(\alpha)}{\displaystyle\partial
\phi}=exp(-\alpha) [-\frac{1}{2}\dot{\alpha}\partial_{0}\phi
\nonumber
\\ +(\frac{1}{4}\dot{\alpha}^{2}-\mu^{2})\phi
+exp(-\alpha)\lambda\phi^{3}]=0
\\
\frac{\displaystyle\partial H(\alpha)}{\displaystyle\partial
\alpha}=-exp(-\alpha)
[-\frac{1}{2}\dot{\alpha}\phi\partial_{0}\phi
+\frac{1}{2}\sum_{j=0}^3 (\partial_{j}\phi)^{2}+ \nonumber
\\ (\frac{1}{8}\dot{\alpha}^{2}-\frac{1}{2}\mu^{2})\phi^{2}
+\frac{1}{2}exp(-\alpha)\lambda\phi^{4}]=0
\end{array}\right.
\end{equation}
That is equivalent to
\begin{equation}\label{U3.8I}
\left\{
\begin{array}{ll}
\frac{\displaystyle\partial H(\alpha)}{\displaystyle\partial
\phi}\sim[-\frac{1}{2}\dot{\alpha}\partial_{0}\phi
+(\frac{1}{4}\dot{\alpha}^{2}-\mu^{2})\phi
+exp(-\alpha)\lambda\phi^{3}]=0
\\
\frac{\displaystyle\partial H(\alpha)}{\displaystyle\partial
\alpha}\sim [-\frac{1}{2}\dot{\alpha}\phi\partial_{0}\phi
+\frac{1}{2}\sum_{j=0}^3 (\partial_{j}\phi)^{2} \nonumber \\
+(\frac{1}{8}\dot{\alpha}^{2}-\frac{1}{2}\mu^{2})\phi^{2}
+\frac{1}{2}exp(-\alpha)\lambda\phi^{4}]=0
\end{array}\right.
\end{equation}
Now consider different solutions for (\ref{U3.7I}) or
(\ref{U3.8I}).
\\
I. In case of deformation parameter $\alpha$ weakly dependent on
time
\begin{equation}\label{U3.9I}
\dot{\alpha}\approx 0 .
\end{equation}
Most often this happens at low energies (far from the Planck's).
\\ Actually, as $\alpha=l_{min}^{2}/a(t)^{2}$,
where $a(t)$ is the measuring scale, (\ref{U3.9I}) is nothing else
but $\alpha \approx 0$.  And the process takes place at low
energies or at rather high energies but at the same energy scale,
meaning that scale factor  $a(t)^{-2}$ is weakly dependent on
time. The first case is no doubt more real. In both cases,
however, we have a symmetry breakdown and for minimum of
$\tilde{\sigma}$ in case under consideration
\begin{equation}\label{U3.10I}
\tilde{\sigma} = \pm\mu\lambda^{-1/2}exp(\alpha/2).
\end{equation}
Based on the results of \cite{shalyt5},\cite{shalyt14}, it follows
that
\begin{equation}\label{U3.11I}
<0\mid \phi\mid0>_{\alpha}=exp(-\alpha)<0\mid \phi \mid0>,
\end{equation}
and we directly obtain
\begin{equation}\label{U3.11AI}
<0\mid\tilde{\sigma}\mid0>_{\alpha}=\pm\mu\lambda^{-1/2}exp(-\alpha/2)..
\end{equation}
Then in accordance with the exact formula \cite{cosm3},\cite{Div1}
we shift the field $\phi(\alpha)$
\begin{equation}\label{U3.12I}
\phi(\alpha) \Rightarrow \phi(\alpha)+<0\mid
\tilde{\sigma}\mid0>_{\alpha}=exp(-\alpha/2)(\phi+\sigma)=
\phi(\alpha)+\sigma(\alpha),
\end{equation}
where $\sigma=\pm\mu\lambda^{-1/2}$- minimum in a conventional
nondeformed case \cite{cosm3}. When considering $\alpha$ -
deformed Lagrangian (\ref{U3.3I}) for the shifted field
$\phi+\sigma$
\begin{equation}\label{U3.13I}
L(\alpha)=L(\alpha,\phi+\sigma),
\end{equation}
we obtain as expected a massive particle the squared mass of that
contains, compared to a well-known case, the multiplicative
exponential supplement $exp(-\alpha)$
\begin{equation}\label{U3.14I}
m_{\phi}^{2}=2\mu^{2}exp(-\alpha),
\end{equation}
leading to the familiar result for low energies or correcting the
particle's mass in the direction of decreasing values for high
energies.
\\
II. Case of $\dot{\alpha} \neq 0$.
\\ This case is associated with a change from low to higher energies
or vice versa. This case necessitates the following additional
assumption:
\begin{equation}\label{U3.15I}
\partial_{0}\phi\approx 0.
\end{equation}
What is the actual meaning of the assumption in (\ref{U3.15I})? It
is quite understandable that in this case in $\alpha$ - deformed
field $\phi(\alpha)=exp(-\frac{\alpha(t)}{2})\phi$ the principal
dependence on time $t$ is absorbed by exponential factor
$exp(-\frac{\alpha(t)}{2})$. This is quite natural at sufficiently
rapid changes of the scale (conforming to the energy) that is just
the case in situation under study.
\\ Note that in the process a change from Lagrangian to the
Hamiltonian formalism (\ref{U3.4I}) for $\alpha$ - deformed
Lagrangian $L(\alpha)$ (\ref{U3.3I}) holds true since, despite the
condition of (\ref{U3.15I}), we come to
\begin{equation}\label{U3.16I}
\partial_{0}\phi(\alpha)\neq 0
\end{equation}
due to the presence of exponential factor
$exp(-\frac{\alpha(t)}{2})$.
\\Proceed to the solution for (\ref{U3.7I}) (and respectively (\ref{U3.8I})).
Taking (\ref{U3.15I}) into consideration, for a minimum of
$\tilde{\sigma}$ in this case we obtain
\begin{equation}\label{U3.17I}
\tilde{\sigma} =
\pm(\mu^{2}-\frac{1}{4}\dot{\alpha}^{2})^{1/2}\lambda^{-1/2}exp(\alpha/2).
\end{equation}
So, the requisite for the derivation of this minimum will be as
follows:
\begin{equation}\label{U3.18I}
4\mu^{2}-\dot{\alpha}^{2}\geq 0
\end{equation}
or with the assumption of $\mu>0$
\begin{equation}\label{U3.19I}
-2\mu \leq\dot{\alpha} \leq 2\mu.
\end{equation}
Assuming that $\alpha(t)$ is increasing in time, i.e. on going
from low to higher energies and with $\dot{\alpha}>0$, we have
\begin{equation}\label{U3.20I}
 0<\dot{\alpha}\leq 2\mu
\end{equation}
or\begin{equation}\label{U3.21I}
\alpha(t) \leq 2\mu t.
\end{equation}
From where it follows that on going from low to higher energies,
i.e. with increasing energy of model (\ref{U3.3I}), two different
cases should be considered.
\\
IIa. Symmetry breakdown when $\alpha(t) < 2\mu t$
\\
$$\tilde{\sigma} =
\pm(\mu^{2}-\frac{1}{4}\dot{\alpha}^{2})^{1/2}\lambda^{-1/2}exp(\alpha/2).
$$
\\
IIb. Symmetry restoration when $\alpha(t) = 2\mu t$ as in this
case
\\
$$\tilde{\sigma}=0.$$
\\
Because of $\alpha(t) \sim a(t)^{-2}$, these two cases IIa and IIb
may be interpreted as follows: provided $a(t)$ increases more
rapidly than $(2\mu t)^{-1/2}$ (to within a familiar factor), we
have a symmetry breakdown at hand as in the conventional case
(\ref{U3.1I}), whereas for similar increase a symmetry restoration
occurs. This means that at sufficiently high energies associated
with scale factor $a(t)\sim (2\mu t)^{-1/2}$ the broken symmetry
is restored.
\\ Provided that in this case the time dependence of $a(t)$
is exactly known, then by setting the equality
\\
$$a(t)=l_{min}(2\mu t)^{-1/2}$$
\\
and solving the above equation we can find $t_{c}$ - critical time
for the symmetry restoration. Then the critical scale (critical
energies)
\\
$$a(t_{c})=l_{min}(2\mu t_{c})^{-1/2},$$
\\
actually the energies whose symmetry is restored, and finally the
corresponding critical point of the deformation parameter
\\
$$\alpha(t_{c})=l_{min}^{2}a(t_{c})^{-2}=2\mu t_{c}.$$
\\
This point is critical in a sense that for all points with the
deformation parameter below its critical value
\\
$$\alpha(t)<\alpha(t_{c})$$
\\
the symmetry breakdown will be observed. Obviously, in case under
study the energy is constantly growing with corresponding lowering
of the scale and hence
\\
$$a(t)\sim t^{\xi},\xi<0.$$
\\
\\ Of particular interest is a change from high to lower energies.
This is associated with the fact that all cosmological models may
be involved, i.e. all the cases where scale $a(t)$ is increased
due to the Big Bang \cite{cosm1},\cite{cosm2}.
\\ For such a change with the assumption that $\alpha(t)$ diminishes
in time and $\dot{\alpha}<0$ we obtain
\begin{equation}\label{U3.22I}
-2\mu \leq\dot{\alpha}<0.
\end{equation}
Since by definition $\alpha(t)>0$ is always the case and
considering $\alpha(t)$ as a negative increment (i.e.
$d\alpha(t)<0$), we come to the conclusion that the case under
study is symmetric to IIa and IIb.
\\IIc.  For fairly high energies, i.e. for $\alpha(t) = 2\mu t$
or scale $a(t)$ that equals to $(2\mu t)^{-1/2}$(again to within
the familiar factor), there is no symmetry breakdown in accordance
with case IIb.
\\
\\IId. On going to lower energies associated with $\alpha(t) < 2\mu t$
there is a symmetry breakdown in accordance with case IIa and with
the formula of (\ref{U3.17I}). Note that in this case energy is
lowered in time and hence the scale is growing, respectively.
Because of this,
\\
$$a(t)\sim t^{\xi},\xi>0.$$
\\
For the specific cases with exactly known relation between $a(t)$
and $t$ one can determine the points without the symmetry
breakdown. In cosmology \cite{cosm2} in particular we have
\\
\\1) in a Universe dominated by nonrelativistic matter
\\
$$a(t)\propto t^{2/3},$$
\\
i.e.
\\
$$a(t)\approx a_{1}t^{2/3}.$$
\\
From whence at the point of unbroken symmetry
\begin{equation}\label{U3.23I}
a_{1}t^{2/3}\approx(2\mu t)^{-1/2},
\end{equation}
directly giving the critical time $t_{c}$
\begin{equation}\label{U3.24I}
t_{c}\approx(2\mu)^{-3/7}a_{1}^{-6/7},
\end{equation}
and for $t>t_{c}$ the symmetry breakdown is observed (case IId).
\\ In much the same manner
\\
\\2) in the Universe dominated by radiation
\\
$$a(t)\propto t^{1/2},$$
\\
i.e.
\\
$$a(t)\approx a_{2}t^{1/2}$$
\\
from where for $t_{c}$ we have
\begin{equation}\label{U3.25I}
t_{c}\approx (2\mu)^{-1/2}a_{2}^{-1},
\end{equation}
and again for $t>t_{c}$ the symmetry is broken.
\\ Here it is interesting to note that despite the apparent symmetry
of cases IIa,IIb and IId , IIc, there is one important
distinction.
\\ In cases IId and IIc time $t$ is usually (in cosmological
models as well \cite{cosm1},\cite{cosm2}) counted from the Big
Bang moment and therefore fits well to the associated time
(temperature) coordinate of the lattice
$Lat^{\tau}_{\widetilde{\alpha}}$ (section 8 and
\cite{shalyt14},\cite{shalyt7}). As a result, when the critical
time $t_{c}$ is known, one can find the critical point of the
above-mentioned lattice as follows:
$(\alpha_{c},\tau_{c})=(\widetilde{\alpha_{c}},\tau_{c})$, where
all the three coordinates of the space part
$Lat^{\tau}_{\widetilde{\alpha}}$, i.e. in
$(\widetilde{\alpha_{c}})$, are equal to
\\
\\$$\alpha_{c}=l_{min}^{2}a(t_{c})^{-2},$$
\\
and
\\
\\$$\tau_{c}=T_{c}^{2}/T_{max}^{2},$$
\\
where $T_{c}\sim 1/t_{c}$ ((\ref{U15t}) in section 6 and
\cite{shalyt7},\cite{shalyt9},\cite{shalyt10}).
\\ Thus, in these cases all points $Lat^{\tau}_{\widetilde{\alpha}}$,
for which the following conditions are satisfied:
\begin{equation}\label{U3.26I}
\left\{
\begin{array}{ll}
\alpha<\alpha_{c},
\\
\tau<\tau_{c}
\end{array}\right.
\end{equation}
are associated with a symmetry breakdown, whereas at the critical
point $(\alpha_{c},\tau_{c})$ no symmetry breakdown occurs. As
seen from all the above formulae, the point of the retained
symmetry is associated with higher temperatures and energies than
those (\ref{U3.26I}), where a symmetry breakdown takes place, in
qualitative agreement with the principal results of \cite{cosm3}.
\\ Note that for cases IIa and IIb time $t$ is a certain local time
of the quantum process having no direct relation to the time
(temperature) variable of lattice
$Lat^{\tau}_{\widetilde{\alpha}}$. Because of this, it is possible
to consider only the critical value at the space part
$Lat_{\widetilde{\alpha}}$ \cite{shalyt14},\cite{shalyt15} of
lattice $Lat^{\tau}_{\widetilde{\alpha}}$, i.e. $\alpha_{c}\in
Lat_{\widetilde{\alpha}}$, all other inferences remaining true
with a change of (\ref{U3.26I}) by
\begin{equation}\label{U3.26AI}
\alpha<\alpha_{c}.
\end{equation}
However, provided there is some way to find temperature $T_{c}$
that is associated with a symmetry restoration(e.g, \cite{cosm3}
case IIb), one can directly calculate $\tau_{c}$ and finally
change (\ref{U3.26AI})by (\ref{U3.26I}).
\\ In any case conditions $0<\alpha\leq 1/4;0<\tau\leq 1/4$
(sections 3,5 and \cite{shalyt4},
\cite{shalyt5},\cite{shalyt7},\cite{shalyt14},\cite{shalyt15})
impose constraints on the model parameters (\ref{U3.26I}) and
hence on $\mu$, which may be found in the explicit form by solving
the following inequality:
\\
\\$$0<\alpha_{c}\leq 1/4.$$
\\Thus, from primary analysis of such a simple model as (\ref{U3.1I})
it is seen that its $\alpha$ - deformation (\ref{U3.3I})
contributes considerably to widening the scope of possibilities
for a symmetry breakdown and restoration.
\\
\\1) At low energies (far from the Planck's) it reproduces with
a high accuracy (up to $exp(-\alpha)\approx 1$)  the results
analogous to those given by a nondeformed theory \cite{cosm3}.
\\
\\2) On going from low to higher energies or vice versa it provides
new cases of a symmetry breakdown or restoration depending on the
variation rate of the deformation parameter in time, being capable
to point to the critical points of the earlier considered lattice
$Lat^{\tau}_{\widetilde{\alpha}}$ , i.e. points of the symmetry
restoration.
\\
\\3) This model makes it possible to find important constraints
on the parameters of the initial model having an explicit physical
meaning.
\\ It should be noted that for $\alpha$ - deformed theory (\ref{U3.3I})
 there are two reasons to be finite in ultraviolet:
\\ - cut-off for a maximum momentum $p_{max}$;
\\ - damping factors of the form $exp(-\alpha)$,
where in the momentum representation we have
$\alpha=p^{2}/p_{max}^{2}$ in each order of a perturbation theory
are suppressing the greatest momenta. This aspect has been already
touched upon in section 3.2.
\\ As a result, some problems associated with a conventional
theory \cite{cosm3} (divergence and so on) in this case are
nonexistent.

\section{Conclusion}
In conclusion the scope of problems associated with the
above-mentioned methods is briefly outlined.
\\
{\bf \\I.Involvement of Heisenberg's Algebra Deformation}
\\ One of the major problems associated with the proposed
approach to investigation of Quantum Mechanics of the Early
Universe is an understanding of its relation to the Heisenberg' s
algebra deformation(e.g. see \cite{Magg}). It should be noted that
from the author's point of view the latter has two serious
disadvantages:
\\{\bf 1)} the deformation parameter is a dimensional
variable $\kappa$ with a dimension of mass;
\\{\bf 2)} in the limiting transition to QM this parameter goes
to infinity and fluctuations of other values are hardly sensitive
to it.
\\ At the same time, the merit of this approach is its ability
with particular assumptions to reproduce the Generalized
Uncertainty Relations.
\\ The proposed approach is free from such limitations as
{\bf 1)} and {\bf 2)}, since the deformation parameter is
represented by the dimensionless quantity $\alpha$ and the
variation interval $\alpha$ is finite $0<\alpha\leq1/4$. However,
it provides no direct reproduction of the Generalized Uncertainty
Relations. This approach is applicable in the general cases of
Quantum Mechanics with Fundamental Length irrespective of the fact
whether it is derived from the Generalized Uncertainty Relations
or in some other way.
\\ It should be noted that a present-day approach to the description
of Quantum Mechanics at Planck's scales is termed more generally:
Lie-algebraic deformation. There is a reason to
believe that the approach developed by the author in the
above-mentioned papers is consistent with the research trend
associated with Lie-algebraic deformation
\cite{r5},\cite{Magg},\cite{r29},\cite{Chrys}, \cite{Mend},
probably being its extension. At first, the author's interest has
been focused at possible variations in the measurement procedure
with the appearance of the fundamental length (i.e. close to the
space-time singularities, where the quantum gravitational effects
are significant) rather than at kinematics \cite{r1}). Because of
this, the density matrix and average values of the operators were
subjected to deformation as the principal objects. In this
approach the chosen deformation parameter $\alpha$
 has an explicit physical meaning: it indicates an extent to
which  the energies under study  are differing from a maximum
energy. The main inferences from this approach are concerned with
a quantum theory of black holes
\cite{shalyt1},\cite{shalyt5},\cite{shalyt11},\cite{shalyt13},
\cite{shalyt13s} extension of the entropy concept in particular.
So, the author's motivation in the development of this approach is
elucidation of the inferences introduced into a quantum theory of
black holes with the involvement of the fundamental length. As
noted above, these inferences are quite numerous and they are
fairly adequate within the present day paradigm, extending it
without apparent conflict. Of course, compared to
\cite{r5},\cite{Magg},\cite{r29},\cite{Chrys}, \cite{Mend}, in the
proposed approach (density matrix deformation) noncommutativity of
the space coordinates is not used in principle. At the same time,
this feature is not inconsistent with the proposed approach. In
this way the approach necessitates further investigation and
extension. Unfortunately, the author is still in search for the
algebraic base of the density-matrix deformation approach that
could be used in description of quantum mechanics of the Early
Universe, since his interest so far has been concerned primarily
with various physical inferences and applications of the approach.
It should be noted that these inferences are similar for different
approaches: black mini-holes with a mass on the order of Planck's
$\sim M_{p}$ are stable both in the density matrix deformation
approach \cite{shalyt5}) and Generalized Uncertainty Relations
approach \cite{r17}.
\\Because of this, involvement of the both approaches
in deformation of Quantum Mechanics is of particular importance.
\\{\bf \\II. The Approach as Applied to a Quantum Theory of Black Holes}
\\
{\bf 2.1 Bekenstein-Hawking formula strong derivation}
\\This paper presents certain results pertinent to the application of the
above methods in a Quantum Theory of Black Holes (subsections {\bf
4.2, 4.3}). Further investigations are still required in this
respect, specifically for the complete derivation of a
semiclassical Bekenstein-Hawking formula for the Black Hole
entropy, since in subsection 4.3 the treatment has been based on
the demonstrated result: a respective number of the degrees of
freedom is equal to $A$, where $A$ is the surface area of a black
hole measured in Planck's units of area $L_{p}^{2}$
(e.g.\cite{r14},\cite{r20}). Also it is essential to derive this
result from the basic principles given in this paper. Problems
{\bf 2.1} and {\bf 2.2} are related.
\\{\bf 2.2  Calculation of quantum corrections to the Bekenstein-Hawking
 formula}.
\\ In subsection 8.4 for the introduced logarithmic correction it
has been noted (see for example \cite{r22}) that it is coincident
with coefficient $a_{0}$  in formula (\ref{U28L}):
\\
$$Sp[\rho(\alpha)]-Sp^{2}[\rho(\alpha)]=\alpha+a_{0}\alpha^{2}
+a_{1}\alpha^{3}+... $$
\\
when using the exponential ansatz. It is clear that such a
coincidence is not accidental and further investigations are
required to elucidate this problem.
\\{\bf 2.3 Quantum mechanics and thermodynamics of black holes with
GUR}
\\Of interest is to consider the results of \cite{r17}, \cite{r31}
as related to the quantum-mechanical studies and thermodynamics of
black holes with GUR assumed valid rather than the Heisenberg
Uncertainty Relations. This is directly connected to the
above-mentioned problem of the associations between the density
matrix deformation considered in this work and Heisenberg's
algebra deformation.
\\
\\{\bf 2.4 Singularities and cosmic censorship hypothesis}
\\In subsection {\bf 4.2} a slight recourse has been made to the
case when Schwarzshild radius is $r=0$ that is associated with
going to value $\alpha=1$ and finally to a complex value of the
density pro-matrix trace $Sp[\rho (\alpha)]$. It should be noted
that the problem of singularities is much more complex \cite{r12},
\cite{Pen1},\cite{Sing} and is presently treated both physically
and mathematically. It seems interesting to establish the
involvement of the results obtained by the author in solving of
this problem.
\\
\\{\bf III. Divergence in Quantum Field Theory}
\\It is obvious that once the fundamental length is included into a
Quantum Theory, ultra-violet (UV) divergences should be excluded
due to the presence of a maximum momentum determining the cut-off
\cite{r1}. In case under study this is indicated by the presence
of an exponential ansatz (subsection {\bf 3.2}). Note, however,
that for any particular theory it is essential to derive the
results from the basic principles with high accuracy and in good
agreement with the already available ones and with the
experimental data of QFT for the UV region without renormalization
\cite{Div1,Div2}.
\\{\bf IV. The Approach as Applied to Inflation Cosmology}
As we concern ourselves with the Early Universe (Planck's
energies), the proposed methods may be applied in studies of
inflation cosmology \cite{r10,r27,Inf,cosm2}, especially as
Wheeler-DeWitt Wave Function of the Universe $\Psi$ \cite{r11} is
reliably applicable in a semiclassical approximation only
\cite{Vil}. The problem is formulated as follows: on what
conditions and in what way the density pro-matrix $\rho (\alpha)$
or its respective modification may be a substitute for $\Psi$ in
inflation models?
\\ Besides, it is interesting to find, to what changes may be subjected
the cosmological perturbations attached to inflation according to
the deformation of Quantum Mechanics considered in the present
paper. It is necessary to elucidate possible changes in the
quantum-mechanical origin \cite{J} and their inferences directly
applicable to the inflation theory
\\{\bf V. High-Energy Deformation of Gravitation}
\\Since this work actually presents a study of physics at Planck's
scales, it is expedient to consider quantum-gravitational effects
which should be incorporated for specific energies. As a
development of the proposed approach this means the construction
of an adequate deformation of the General Relativity including
parameter $\alpha$, i.e. deformation of Einstein's Equations and
the associated Lagrangian involving parameter $\alpha$. Then the
question arises: and what about the space-time quantization? The
author holds the viewpoint that as the first approximation of a
quantized space-time one can use a portion of the Nonuniform
Lattice $Lat^{\tau}_{\widetilde{\alpha}}$ described in section 8
that is associated with small-number nodes or with high-valued
parameters $\widetilde{\alpha}$ and $\tau$ ,just which are used to
define the physics at Planck scale where the quantum-gravitational
effects are considerable. In this approximation for the prototype
of a point in the General Relativity may be taken an elementary
cell, i.e. as an element of the above-mentioned lattice with
small-number neighboring nodes. Then the associated deformation of
Einstein's should be considered exactly in this cell.
\\ Note that this section involves all the problems considered in I-IV.
\\ In summary it might be well to make three general remarks.
\\
\\{\bf 1)} It should be noted that in some well-known papers on GUR and
Quantum Gravity (e.g. see \cite{r1,r3,r5,Magg}) there is no
mention of any measuring procedure. However, it is clear that this
question is crucial and it cannot be ignored or passed over in
silence. We would like to remark that the measuring rule used in
\cite{r29}, (formula (5)) is identical to the ours.  In this paper
the proposed measuring rule (\ref{U6}) is a good initial
approximation to the exact measuring procedure of QMFL.
Corrections to this procedure could be defined by an adequate and
fully established description of the space-time foam (see
\cite{r25},\cite{Foam}) at Planck's scale.
\\
\\{\bf 2)} One of the principal issues of the present work is the
development of a unified approach to study all the available
quantum theories without exception owing to the proposed small
dimensionless deformation parameter $\widetilde{\alpha}_{\tau}\in
Lat^{\tau}_{\widetilde{\alpha}}$ that is in turn dependent on all
the fundamental constants $G,c,\hbar$ and  $k$.
\\ Thus, there is  reason to believe that lattices
$Lat_{\widetilde{\alpha}}$ and $Lat^{\tau}_{\widetilde{\alpha}}$
may be a universal means to study different quantum theories. This
poses a number of intriguing problems:
\\{\bf (1)} description of a set of lattice symmetries
$Lat_{\widetilde{\alpha}}$ and $Lat^{\tau}_{\widetilde{\alpha}}$;
\\ {\bf (2)} for each of the well-known physical theories
($\varphi^{4}$,QED,QCD and so on) definition of the selected
(special) points (phase transitions, different symmetry violations
and so on) associated with the above-mentioned lattices.
\\{\bf 3)} As it was noted in \cite{Fadd},
advancement of a new physical theory implies the introduction of a
new parameter and deformation of the precedent theory by this
parameter. In essence, all these deformation parameters are
fundamental constants: $G$, $c$ and $\hbar$ (more exactly in
\cite{Fadd} $1/c$ is used instead of $c$). As follows from the
above results, in the problem from \cite{Fadd} one may
redetermine, whether a theory we are seeking is the theory with
the fundamental length involving these three parameters by
definition: $L_{p}=\sqrt\frac{G\hbar}{c^3}$. Notice also that the
deformation introduced in this paper is stable in the sense
indicated in \cite{Fadd}.
\\
\\{\bf Acknowledgements}
\\The author would like to acknowledge Prof. Nikolai Shumeiko,
Director of the Belarusian National Center of Particles and High-
Energy Physics, and Dr. Julia  Fedotova,Scientific Secretary of
the Center, for their assistance contributing to realization of my
research plans and activities; Ludmila Kovalenko for her
assistance in editing and Sofia Titovich for her help in
preparation of this manuscript; Prof.D.V.Ahluwalia-Khalilova,
Center for Studies of Physical, Mathematical and Biological
Structure of Universe, Department of Mathematics, University of
Zacatecas, Mexico for his interest in the subject matter; Profs.
Sergei Kilin, Vassilii Strazhev and also Dr. Arthur Tregubovich
for valuable discussions and remarks; and last but not the least
my wife Nadya Anosova for the support and encouragement when
working on this chapter.


\end{document}